\documentclass{article}
\usepackage{authblk}

\title{\bslack s: Highly Space Efficient B-trees}

\author{Trevor Brown}
\affil{Institute of Science and Technology, Austria\\me@tbrown.pro}

\usepackage[usenames,dvipsnames,svgnames,table]{xcolor}
\definecolor{trevcolor}{RGB}{0,80,200}
\definecolor{pdfbgcolor}{RGB}{180,180,180}

\newcommand{\later}[1]{}
\newcommand{\trevorlater}[1]{}

\usepackage{hyperref}
\hypersetup{linktocpage}

\usepackage{etoolbox}
\usepackage{environ}
\newtoggle{isthesis}
\togglefalse{isthesis}
\NewEnviron{thesisonly}{%
  \iftoggle{isthesis}
    {\BODY\xspace}
    {}%
  }
\NewEnviron{thesisnot}{%
  \iftoggle{isthesis}
    {}
    {\BODY\xspace}%
  }

\NewEnviron{fakethm}{%
  \medskip\noindent\textbf{Theorem}\quad %
  \textit{\BODY}%
  \medskip%
}
\NewEnviron{fakeinv}{%
  \medskip\noindent\textbf{Invariant}\quad %
  \textit{\BODY}%
  \medskip%
}
\NewEnviron{fakeprop}{%
  \medskip\noindent\textbf{Proposition}\quad %
  \textit{\BODY}%
  \medskip%
}
\NewEnviron{fakelem}{%
  \medskip\noindent\textbf{Lemma}\quad %
  \textit{\BODY}%
  \medskip%
}
\NewEnviron{fakedefn}{%
  \medskip\noindent\textbf{Definition}\quad %
  \textit{\BODY}%
  \medskip%
}

\usepackage{framed}
\usepackage{tabto}
\usepackage{comment}
\excludecomment{ignore}
\usepackage{amsfonts}
\usepackage{amssymb}
\usepackage{graphicx}
\usepackage{setspace}
\usepackage{smartref} 
\usepackage{fullpage}
\usepackage{listings}
\usepackage{ragged2e}
\usepackage{enumerate}
\usepackage{rotating}
\usepackage{verbatim}
\usepackage[normalem]{ulem}
\usepackage{paralist}
\usepackage{wrapfig}
\usepackage{paralist}
\usepackage{tabularx}
\usepackage[shortlabels]{enumitem}

\usepackage{pgf}

\newcolumntype{Y}{>{\centering\arraybackslash} X}

\newcommand{\qed}{\hfill{\rule{2mm}{2mm}}}
\newenvironment{chapscxproof}{\begin{trivlist}
\item[\hspace{\labelsep}{\bf\noindent Proof: }]
}{\qed\end{trivlist}}

\newcommand{\func}[1]{\mbox{\sc #1}}

\newcommand{\nil}{\textsc{Nil}}

\lstdefinestyle{nonumbers}{numbers=none}

\newtheorem{thm}{Theorem}

\newtheorem{lem}[thm]{Lemma}
\newtheorem{cor}[thm]{Corollary}

\newtheorem{prop}[thm]{Proposition}

\newcommand{\ourcomments}[2]{{\bf [[[#1--#2]]]}}

\newcommand{\trevor}[1]{\ourcomments{#1}{Trevor}}

\newcommand{\after}[1]{}

\newcommand{\maxslack}{overslack tree}
\newcommand{\bslack}{B-slack tree}
\newcommand{\rbslack}{relaxed \bslack}
\newcommand{\Rbslack}{Relaxed \bslack}
\newcommand{\weight}{weight}

\newcommand{\details}[1]{}

\newcommand{\iset}{A-multiset}

\usepackage[mathcal]{euscript}

\usepackage{tikz}

\usepackage[T1]{fontenc}
\usepackage[utf8]{inputenc}
\usepackage{mathptmx}
\usepackage{pifont}

\lstdefinestyle{customc}{
  belowcaptionskip=1\baselineskip,
  breaklines=true,
  frame=L,
  xleftmargin=\parindent,
  language=C++,
  showstringspaces=false,
  basicstyle=\scriptsize\ttfamily,
  keywordstyle=\bfseries\color{green!40!black},
  commentstyle=\itshape\color{purple!40!black},
  identifierstyle=\color{blue},
  stringstyle=\color{orange},
  numbers=left
}

\usepackage{amsfonts}
\usepackage{amssymb}
\usepackage{epsfig}
\usepackage{subcaption}

\usepackage{amsmath}
\usepackage{listings}
\usepackage{array}
\usepackage{paralist}
\usepackage{url}
\usepackage{xspace}
\usepackage[final]{showlabels}
\usepackage{afterpage}

\usepackage{etoolbox}
\usepackage{environ}
\newtoggle{isfullver}
\NewEnviron{fullver}{%
  \iftoggle{isfullver}
    {\BODY\xspace}
    {}%
  }
\NewEnviron{shortver}{%
  \iftoggle{isfullver}
    {}
    {\BODY\xspace}%
  }
\toggletrue{isfullver}

\usepackage{floatpag}
\floatpagestyle{empty}
\makeatletter
\AtBeginEnvironment{figure}{%
  \@currsize}%
\makeatother
\makeatletter
\AtBeginEnvironment{figure*}{%
  \@currsize}%
\makeatother

\usepackage{array}
\usepackage{xcolor}
\usepackage{graphicx}
\usepackage{rotating}
\usepackage{listings}
\usepackage{paralist}
\usepackage{tikz}
\usepackage{url}
\usepackage{wrapfig}

\usepackage[T1]{fontenc}
\usepackage[utf8]{inputenc}
\usepackage{mathptmx}

\date{}

\begin{document}

\maketitle

\begin{abstract}
\bslack s, a subclass of B-trees that have substantially better worst-case space complexity, are introduced.
They store $n$ keys in height $O(\log_b n)$, where $b$ is the maximum node degree. Updates can be performed in $O(\log_{\frac b 2} n)$ amortized time.
A relaxed balance version, which is well suited for concurrent implementation, is also presented.
\end{abstract}

%

\section{Introduction}

B-trees are balanced trees designed for block-based storage media.
Internal nodes contain between $b/2$ and $b$ child pointers, and one less key.
Leaves contain between $b/2$ and $b$ keys.
All leaves have the same depth, so access times are predictable.
If memory can be allocated on a per-byte basis, nodes can simply be allocated the precise amount of space they need to store their data, and no space is wasted.
However, typically, all nodes have the same, fixed capacity, and some of the capacity of nodes is wasted.
As much as 50\% of the capacity of each node is wasted in the worst case.
This is particularly problematic when data structures are being implemented in hardware, since memory allocation and reclamation schemes are often very simplistic, allowing only a single block size to be allocated (to avoid fragmentation).
Furthermore, since hardware devices must include sufficient resources to handle the worst-case, good expected behaviour is not enough to allow hardware developers to reduce the amount of memory included in their devices.
To address this problem, we introduce \textit{\bslack s}, which are a variant of B-trees with substantially better worst-case space complexity.
We also introduce \rbslack s, which are a variant of \bslack s that are more amenable to concurrent implementation.

The development of \bslack s was inspired by a collaboration with a manufacturer of internet routers, who wanted to build a concurrent router based on a tree.
In such embedded devices, storage is limited, so it is important to use it as efficiently as possible.
A suitable tree would have a simple algorithm for updates, small space complexity, fast searches, and searches that would not be blocked by concurrent updates.
Updates were expected to be
infrequent.
One naive approach is to rebuild the entire tree after each update.
Keeping an old copy of the tree while rebuilding a new copy would allow searches to proceed unhindered, but this would double the space required to store the tree.

Search trees can be either node-oriented, in which each key is stored in an internal node or leaf, or leaf-oriented, in which each key is stored at a leaf and the keys of internal nodes serve only to direct a search to the appropriate leaf.
In a node-oriented B-tree, the leaves and internal nodes have different sizes (because internal nodes contain keys and pointers to children, and leaves contain only keys).
So, if only one block size can be allocated, a significant amount of space is wasted.
Moreover, deletion in a node-oriented tree sometimes requires stealing a key from a successor (or predecessor), which can be in a different part of the tree.
This is a problem for concurrent implementation, since the operation involves a large number of nodes, namely, the nodes on the path between the node and its successor.

\bslack s are leaf-oriented trees with many desirable properties.
The average degree of nodes is high, exceeding $b-2$ for trees of height at least three.
Their space complexity is better than all of their competitors. 
Consider a dictionary implemented by a leaf-oriented search tree, in which, along with each key, a leaf stores a pointer to associated data.
Suppose that each key and each pointer to a child or to data occupies a single word.
Then, $\frac{2b}{b-3}n$ is an upper bound on the number of words needed to store a \bslack\ with $n > b^3$ keys.
For large $b$, this tends to $2n$, which is optimal.
Section~\ref{sec-analysis} gives a more complex upper bound, which is much better when $b$ is small.
\bslack s have logarithmic height, and the number of rebalancing steps performed after a sequence of $m$ updates to a \bslack\ of size $n$ is amortized $O(\log (n+m))$ per update.
Furthermore, the number of rebalancing steps needed to rebalance the tree can be reduced to amortized \textit{constant} per update at the cost of slightly increased space complexity. 

The rest of this paper is organized as follows.
Section~\ref{sec-bslack-related} surveys related work.
Section~\ref{sec-bslack} introduces \bslack s and \rbslack s.
Height, average degree, space complexity and rebalancing costs of \rbslack s (and, hence, of \bslack s) are analyzed in Section~\ref{sec-analysis}.
Section~\ref{sec-constant-rebalancing} describes the variant of \bslack s with amortized constant rebalancing. 
Section~\ref{sec-bslack-space-complexity} compares the worst-case space complexity of \bslack s to competing data structures.
Single-threaded experimental results appear in Section~\ref{sec-bslack-exp}.
Section~\ref{sec-bslack-cleanup} discusses some implementation issues surrounding rebalancing.
Finally, we conclude in Section~\ref{sec-conclusion}.



\section{Related work} \label{sec-bslack-related}

B-trees were initially proposed by Bayer and McCreight in 1970 \cite{bayer1970organization}. Insertion into a full node in a B-tree causes it to split into two nodes, each half full.
Deletion from a half-full node causes it to merge with a neighbour.
Arnow, Tenenbaum and Wu proposed P-trees \cite{arnow1985p}, which enjoy moderate improvements to \textit{average} space complexity over B-trees, but waste 66\% of each node in the worst case.

A number of generalizations of B-trees have been suggested that achieve much less waste if no deletions are performed.
Bayer and McCreight also proposed B*-trees in \cite{bayer1970organization}, which improve upon the worst-case space complexity of B-trees.
At most a third of the capacity of each node in a B*-tree is wasted.
This is achieved by splitting a node only when it and one of its neighbours are both full, replacing these two nodes by three nodes.
K{\"u}spert \cite{kuspert1983} generalized B*-trees to trees where each node contains between $\lfloor \frac{bm}{m+1} \rfloor$ and $b$ pointers or keys, where $m \le b-1$ is a design parameter.
Such a tree behaves just like a B*-tree everywhere except at the leaves.
An insertion into a full leaf causes keys to be shifted among the nearest $m-1$ siblings to make room for the inserted key.
If the $m-1$ nearest siblings are also full, then these $m$ nodes are replaced by $m+1$ nodes which evenly share keys.
Large values of $m$ yield good worst-case space complexity.
However, with large $m$, updates would be complicated and inefficient in a concurrent setting.

Baeza-Yates and Per-\r{a}ke Larson introduced B+trees with partial expansions \cite{baeza1989performance}.
Several node sizes are used, each a multiple of the block size.
An insertion to a full node causes it to expand to the next larger node size.
With three node sizes, at most 33\% of each node can be wasted, and worst-case utilization improves with the number of block sizes used.
However, this technique simply pushes the complexity of the problem onto the memory allocator. Memory allocation is relatively simple for one block size, but it quickly becomes impractical for simple hardware to support larger numbers of block sizes.

Culik, Ottmann and Wood introduced strongly dense multiway trees (SDM-trees) \cite{culik1981dense}.
An SDM-tree is a node-oriented tree in which all leaves have the same depth, and the root contains at least two pointers.
Apart from the root, every internal node $u$ with fewer than $b$ pointers has at least one sibling.
Each sibling of $u$ has $b$ pointers if it is an internal node and $b$ keys if it is a leaf.
Insertion can be done in $O(b^3 + (\log n)^{b-2})$ time.
Deletion is not supported, but the authors mention that the insertion algorithm could be modified to obtain a deletion algorithm, and the time complexity of the resulting algorithm ``would be at most $O(n)$ and at least $O((\log n)^{b-1})$.''
Besides the long running times for each operation (and the lack of better amortized results), the insertion algorithm is very complex and involves many nodes, which makes it poorly suited for hardware implementation.
Furthermore, in a concurrent setting, an extremely large section of the tree would have to be modified atomically, which would severely limit concurrency.

Srinivasan introduced a leaf-oriented B-tree variant called an \textit{Overflow tree} \cite{Srinivasan01011991}.
For each parent of a leaf, its children are divided into one or more groups, and an overflow node is associated with each group.
The tree satisfies the B-tree properties and the additional requirement that each leaf contains at least $b-1-s$ keys, where $s \ge 2$ is a design parameter and $b$ is the maximum degree of nodes.
Inserting a key into a full leaf causes the key to be inserted into the overflow node instead; if the overflow node is full, the entire group is reorganized.
Deleting from a leaf is the same as in a B-tree unless it will cause the leaf to contain too few keys, in which case, a key is taken from the overflow node; if a key cannot be taken from the overflow node, the entire group is reorganized.
Each search must look at an overflow node.
The need to atomically modify and search two places at once makes this data structure poorly suited for concurrent implementation.

Hsuang introduced a class of node-oriented trees called \textit{H-trees} \cite{Huang:1985:HTO:3857.3858}, which are a subclass of B-trees parameterized by $\gamma$ and $\delta$.
These parameters specify a lower bound on the number of grandchildren of each internal node (that has grandchildren), and a lower bound on the number of keys contained in each leaf, respectively.
Larger values of $\delta$ and $\gamma$ yield trees that use memory more efficiently.
When $\delta$ and $\gamma$ are as large as possible, each leaf contains at least $b-3$ keys, and each internal node has zero or at least $\lfloor \frac{b^2+1}{2} \rfloor$ grandchildren.
The paper presents $O(\log n)$ insertion and deletion algorithms for node-oriented H-trees.
The algorithms are very complex and involve many cases.
H-trees have a minimum average degree of approximately $b/\sqrt{2}$ for internal nodes, which is much smaller than the $b-2$ of \bslack s (for trees of height at least three).

Section~\ref{sec-bslack-space-complexity} 
describes families of B-trees, H-trees and Overflow trees which require significantly more space than  \bslack s.

Rosenberg and Snyder introduced \textit{compact B-trees} \cite{Rosenberg1979}, which can be constructed from a set of keys using the minimum number of nodes possible.
No compactness preserving insertion or deletion procedures are known.
The authors suggested using regular B-tree updates and periodically compacting a data structure to improve efficiency.
However, experiments in \cite{arnow1984empirical} showed that starting with a compact B-tree and adding only 1.6\% more keys using standard B-tree operations reduced storage utilization from 99\% to 67\%.
Thus, to maintain reasonable space complexity, the expensive tree rebuilding/compaction algorithm would have to be executed prohibitively often.

An impressive paper by Br{\"o}nnimann et~al. \cite{bronnimann2007putting} presented three ways to transform an arbitrary sequential dictionary into a more space efficient one.
One of these ways will be discussed here; of the other two, one is extremely complex and poorly suited for concurrent hardware implementation, and the other pushes the complexity onto the memory allocator.

This transformation takes any sequential tree data structure and modifies it by replacing each key in the sequential data structure with a \textit{chunk}, which is a group of $b-2$, $b-1$ or $b$ keys, where $b$ is the memory block size.
All chunks in the data structure are also kept in a doubly linked list to facilitate iteration and movement of keys between chunks. 
For instance, a BST would be transformed into a tree in which each node has zero, one or two children, and $b-2$, $b-1$ or $b$ keys.
All keys in chunks in the left subtree of a node $u$ would be smaller than all keys in $u$'s chunk, and all keys in chunks in the right subtree of $u$ would be larger than all keys in $u$'s chunk.
A search for key $k$ behaves the same as in the sequential data structure until it reaches the only chunk that can contain $k$, and searches for $k$ within the chunk.
An insertion first searches for the appropriate chunk, then it inserts the key into this chunk.
Inserting into a full chunk requires shifting the keys of the $b$ closest neighbouring chunks to make room.
If these $b$ chunks are full, then $b$ consecutive keys from these chunks are first placed in a new chunk, and the remaining $b(b-1)$ keys are distributed amongst the original $b$ chunks. 
Deletion is similar.
Each operation in the resulting data structure runs in $O(f(n)+b^2)$ steps, where $f(n)$ is the number of steps taken by the sequential data structure to perform the same operation.

After this transformation, a B-tree with maximum degree $b$ requires $2n+O(n/b)$ words to store $n$ keys and pointers to data.
In the worst-case, each chunk wastes $2/b$ of its space, which is somewhat worse than in \bslack s.
Furthermore, supporting fast searches can introduce significant complexity to the hardware design.
Suppose this transformation is applied to a B-tree.
Then a node in the transformed B-tree contains up to $b-1$ chunks, each of which contains $b-2, b-1$ or $b$ keys, and occupies one block of memory.
Thus in order to determine which child pointer should be followed, a search must load up to $b-1$ blocks.
In order for searches to be fast, hardware must therefore be able to quickly load up to $b-1$ blocks from memory (a total of $\theta(b^2)$ memory words).
This represents a significant challenge for hardware design.

B-trees with relaxed balance have previously been proposed.
%
%
In particular, several of the updates to \bslack s are derivative of updates to the relaxed $(a,b)$-tree of Larsen~\cite{LF95}.
However, limited work has been done on improving the space complexity of relaxed balance data structures.
Larsen and Fagerberg improved the space complexity of the relaxed $(a,b)$-tree \textit{while the tree is out of balance}~\cite{JL01abtrees}, but its worst-case space complexity remains unchanged. 

\section{\bslack s} \label{sec-bslack}

A \bslack\ is a variant of a B-tree. 
Each node stores its keys in sorted order, so binary search can be used to determine which child of an internal node should be visited next by a search, or whether a leaf contains a key.
Let $p_0, p_1, ..., p_m$ be the sequence of pointers contained in an internal node, and $k_1, k_2, ..., k_m$ be its sequence of keys.
For each $1 \le i \le m$, the subtree pointed to by $p_{i-1}$ contains keys strictly smaller than $k_i$, and the subtree pointed to by $p_i$ contains keys greater than \textit{or equal to} $k_i$.
We say that the \textit{degree of an internal node} is the number of non-\nil\ pointers it contains, and the \textit{degree of a leaf} is the number of keys it contains.
This unusual definition of degree simplifies our discussion.
The degree of node $v$ is denoted $deg(v)$.
If the maximum possible degree of a node is $b$, and its degree is $b-x$, then we say it contains $x$ units of \textit{slack} (or simply $x$ \textit{slack}).

\noindent A \bslack\ is a leaf-oriented search tree with maximum degree $b > 4$ in which:
\begin{compactenum}[\hspace{4.1mm}\bfseries P1:]
\item every leaf has the same depth,
\label{prop-bslack-depth}
\item internal nodes contain between 2 and $b$ pointers (and one less key),%
\label{prop-bslack-internal}
\item leaves contain between 0 and $b$ keys, and
\label{prop-bslack-leaf}
\item for each internal node $u$, the total slack contained in the \textit{children} of $u$ is at most $b-1$.
\label{prop-bslack-slack}
\end{compactenum}
P\ref{prop-bslack-slack} is the key property that distinguishes \bslack s from other variants of B-trees. 
It limits the aggregate space wasted by a number of nodes, as opposed to limiting the space wasted by each node.
Alternatively, P\ref{prop-bslack-slack} can be thought of as a lower bound on the sum of the degrees of the children of each internal node.
Formally, for each internal node with children $v_1, v_2, ..., v_l$, $deg(v_1)+deg(v_2)+...+deg(v_l) \ge lb-(b-1) = lb-b+1$.
This interpretation is useful to show that all nodes have large subtrees.
For instance, it 
implies that a node $u$ with two internal children must have at least $b+1$ grandchildren.
If these grandchildren are also internal nodes, we can conclude that 
$u$ must have at least $b^2-b+2$ great grandchildren.

A tree that satisfies P\ref{prop-bslack-depth}, and in which every node has degree $b-1$, is an example of a \bslack.
Another example of a \bslack\ is a tree of height two, where $b$ is even, the root has degree two, its two children have degree $b/2$ and $b/2+1$, respectively, and the grandchildren of the root are leaves with degree $b$, except for two, one in the left subtree of the root, and one in the right subtree, that each have degree one.
This tree contains the smallest number of keys of any \bslack\ of height two.

\subsection{\Rbslack s} \label{sec-rbslack}

A relaxed balance search tree decouples updates that rebalance (or reorganize the keys of) the tree from updates that modify the set of keys stored in the tree~\cite{relaxedbalance}.
The advantages of this decoupling are twofold.
First, updates to a relaxed balance version of a search tree are smaller, so a greater degree of concurrency is possible in a multithreaded setting.
Second, for some applications, it may be useful to temporarily disable rebalancing to allow a large number of updates to be performed quickly, and to gradually rebalance the tree afterwards.

A \rbslack\ is a relaxed balance version of a \bslack\ that has weakened the properties.
A \weight\ of zero or one is associated with each node.
These \weight s serve a purpose similar to the colors red and black in a red-black tree.
We define the \textit{relaxed depth} of a node to be one less than the sum of the \weight s on the path from the root to this node.
A \rbslack\ is a leaf-oriented search tree with maximum degree $b > 4$ in which:
\begin{compactenum}[\hspace{4.1mm}\bfseries P1$'$:]
\setcounter{enumi}{-1}
\item every node with \weight\ zero contains exactly two pointers,
\label{prop-rbslack-weight-zero}
\item every leaf has the same relaxed depth,
\label{prop-rbslack-depth}
\item internal nodes contain between 1 and $b$ pointers (and one less key), and%
\label{prop-rbslack-internal}
\end{compactenum}
\begin{compactenum}[\hspace{4.1mm}\bfseries P1\hspace{1.01mm}:]
\setcounter{enumi}{2}
\item leaves contain between 0 and $b$ keys
\end{compactenum}

To clarify the difference between \bslack s and \rbslack s, we identify several types of \textit{violations} of the \bslack s properties that can be present in a \rbslack.
We say that a \textit{\weight\ violation} occurs at a node with \weight\ zero, a \textit{slack violation} occurs at a node that violates P\ref{prop-bslack-slack}, and a \textit{degree violation} occurs at an internal node with only one child (violating P\ref{prop-bslack-internal}).
Observe that P1 is satisfied in a \rbslack\ with no \weight\ violations.
Likewise, P2 is satisfied in a \rbslack\ with no degree violations, and P4 is satisfied in a \rbslack\ with no slack violations.
Therefore, a \rbslack\ that contains no violations is a \bslack.
Rebalancing steps can be performed to eliminate violations, and gradually transform any \rbslack\ into a \bslack.

\subsection{Updates to \rbslack s}

\begin{figure}[tbph]
\centering
\begin{tabular}{ | m{2.2cm} | >{\centering\arraybackslash} m{13cm} | }
\hline

\textbf{Delete} & \includegraphics[scale=0.85]{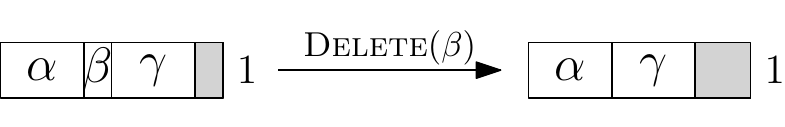} \\
\hline

\vspace{7mm}
\textbf{Insert} & \includegraphics[scale=0.85]{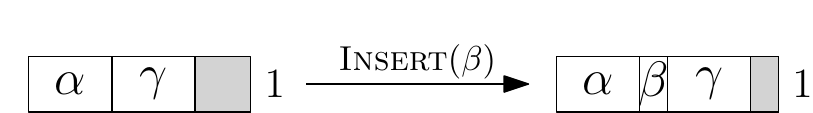} \\
\textbf{Overflow} & \includegraphics[scale=0.85]{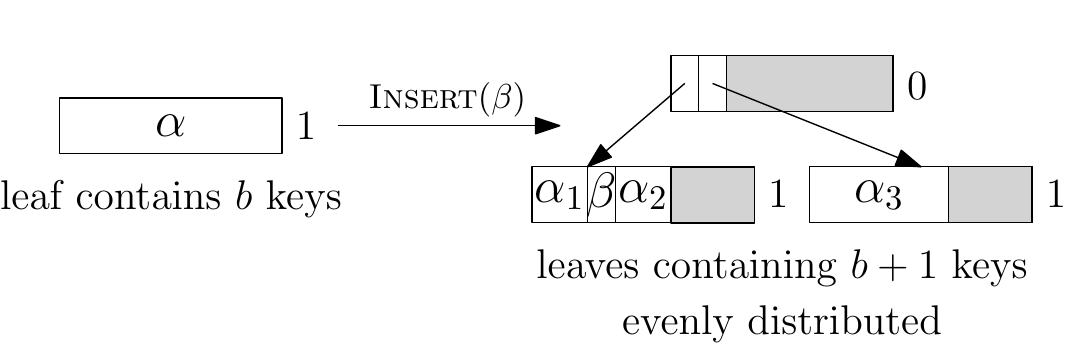} \\
\hline

\vspace{7mm}
\textbf{Root-Zero} & \includegraphics[scale=0.85]{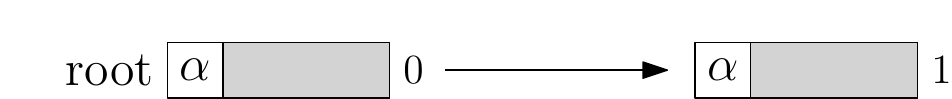} \\
\textbf{Root-Replace} & \includegraphics[scale=0.85]{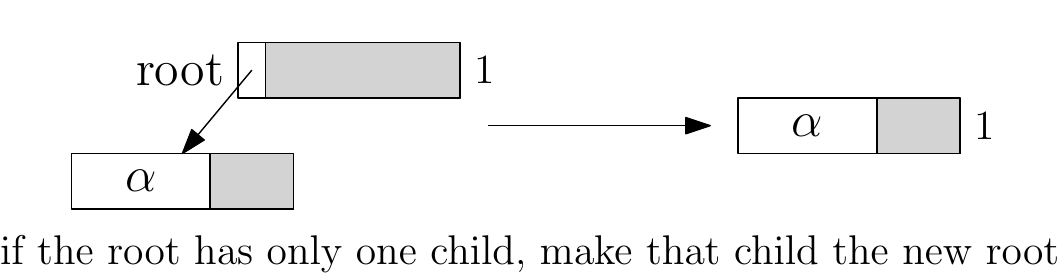} \\
\hline

\vspace{7mm}
\textbf{Absorb} & \includegraphics[scale=0.85]{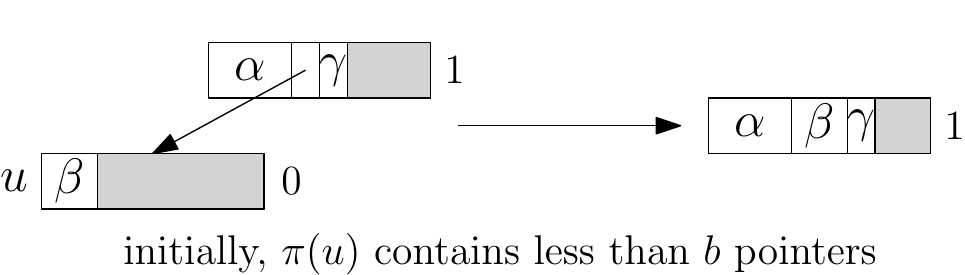} \\
\textbf{Split} & \includegraphics[scale=0.85]{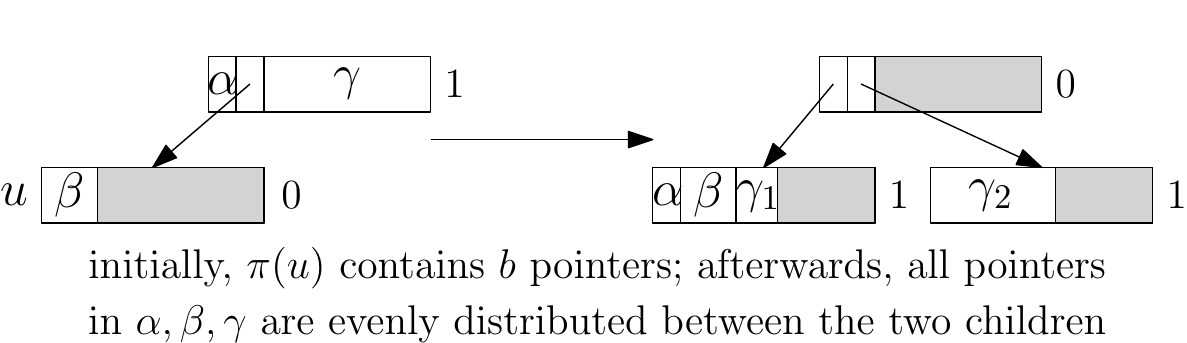} \\
\hline

\vspace{7mm}
\textbf{Compress} & \includegraphics[scale=0.85]{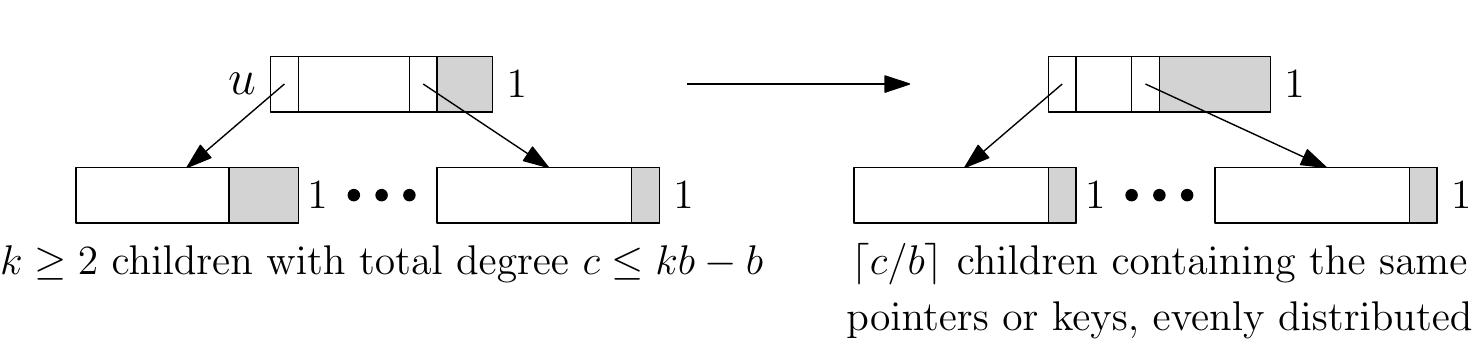} \\
\textbf{One-Child} & \includegraphics[scale=0.85]{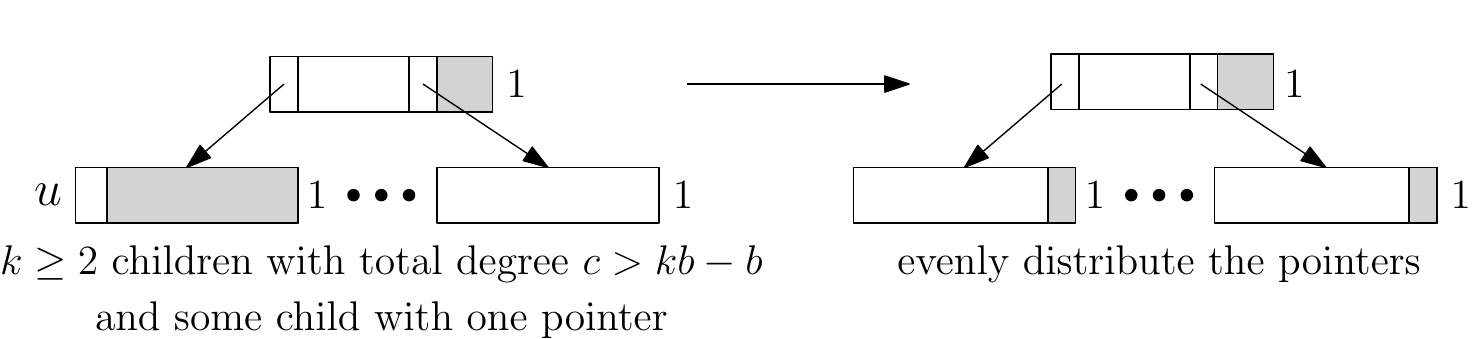} \\
\hline
\end{tabular}
\vspace{-2mm}
\caption{Updates to \bslack s (and \rbslack s).
Nodes with \weight\ zero contain exactly two pointers.
}
\label{fig-bslack-updates}
\end{figure}

We now describe the algorithms for inserting and deleting keys in a \rbslack\ (in a way that maintains P\ref{prop-rbslack-weight-zero}$'$, P\ref{prop-rbslack-depth}$'$, P\ref{prop-rbslack-internal}$'$ and P\ref{prop-bslack-leaf}).
We use the \textbf{Insert}, \textbf{Delete} and \textbf{Overflow} updates introduced by Larsen and Fagerberg in their work on relaxed $(a,b)$-trees~\cite{LF95}.
These updates appear in Figure~\ref{fig-bslack-updates}.
There, \weight s appear to the right of nodes, and shaded regions represent slack.
If $u$ is a node that is not the root, then we let $\pi(u)$ denote the parent of $u$.
The insertion and deletion algorithms always ensure that all leaves have \weight\ one.
We also study how these updates change the amount of slack in nodes, and how they create, move or eliminate violations.

\textbf{Deletion.}
First, a search is performed to find the leaf $u$ where the deletion should occur.
If the leaf does not contain the key to be deleted, then the deletion terminates immediately, and the tree does not change.
If the leaf contains the key to be deleted, then the key is removed from the sequence of keys stored in that leaf. 
Deleting this key may create a slack violation. 


\textbf{Insertion.}
To perform an insertion, a search is first performed to find the leaf $u$ where the insertion should occur.
If $u$ contains some slack, then the key is added to the sequence of keys in $u$, and the insertion terminates.
Otherwise, $u$ cannot accommodate the new key, so Overflow is performed.
Overflow replaces $u$ by a subtree of height one consisting of an internal node with \weight\ zero, and two leaves with \weight\ one.
The $b$ keys stored in $u$, plus the new key, are evenly distributed between the children of the new internal node.
If $u$ was the root before the insertion, then the new internal node becomes the new root.
Otherwise, $u$'s parent $\pi(u)$ before the insertion is changed to point to the new internal node instead of $u$.
After Overflow, there is a \weight\ violation at the new internal node.
Additionally, since the new internal node contains $b-2$ slack, whereas $u$ contained no slack, there may be a slack violation at $\pi(u)$. 
%

Delete, Insert and Overflow maintain the properties of a \rbslack.
They will also 
maintain the properties of a \bslack, provided that rebalancing steps are performed to remove any violations that are created. 

\subsection{Rebalancing steps} \label{sec-rebalancing}

The rebalancing steps are also based on the work of Larsen and Fagerberg.
In fact, \textbf{Root-Zero}, \textbf{Root-Replace}, \textbf{Absorb} and \textbf{Split} are the same as in~\cite{LF95}.
However, the \textbf{Compress} and \textbf{One-Child} operations are newly introduced by this work.
These operations ensure that P\ref{prop-bslack-slack} and P\ref{prop-bslack-internal} are satisfied, respectively.

If there is a degree violation at the root, then Root-Replace is performed.
If there is no degree violation at the root, but there is a weight violation at the root, then Root-Zero is performed.
If there is a \weight\ violation at an internal node that is not the root, then Absorb or Split is performed.
Suppose there are no \weight\ violations.
If there is a degree violation at a node $u$ and no degree or slack violation at $\pi(u)$, then One-Child is performed.
If there is a slack violation at a node $u$ and no degree violation at $u$, then Compress is performed.
Figure~\ref{fig-bslack-updates} illustrates these rebalancing steps.
The goal of rebalancing is to eliminate all violations, while maintaining the \rbslack\ properties.

\textbf{Root-Zero.}
Root-Zero changes the \weight\ of the root from zero to one, eliminating a \weight\ violation, and incrementing the relaxed depth of every node.
If P\ref{prop-bslack-depth}$'$ held before Root-Zero, it holds afterwards.

\textbf{Root-Replace.}
Root-Replace replaces the root $r$ by its only child $u$, and sets $u$'s \weight\ to one.
This eliminates a degree violation at $r$, and any \weight\ violation at $u$.
If $u$ had \weight\ zero before Root-Replace, then the relaxed depth of every leaf is the same before and after Root-Replace.
Otherwise, the relaxed depth of every leaf is decremented by Root-Replace.
In both cases, if P\ref{prop-bslack-depth}$'$ held before Root-Replace, it holds afterwards.

\textbf{Absorb.}
Let $u$ be a non-root node with \weight\ zero.
Absorb is performed when $\pi(u)$ contains less than $b$ pointers.
In this case, the two pointers in $u$ are moved into $\pi(u)$, and $u$ is removed from the tree.
Since the pointer from $\pi(u)$ to $u$ is no longer needed once $u$ is removed, $\pi(u)$ now contains at most $b$ pointers.
The only node that was removed is $u$ and, since it had \weight\ zero, the relaxed depth of every leaf remains the same.
Thus, if P\ref{prop-bslack-depth}$'$ held before Absorb, it also holds afterwards.
Absorb eliminates a \weight\ violation at $u$, but may create a slack violation at $\pi(u)$.

\textbf{Split.}
Let $u$ be a non-root node with \weight\ zero.
Split is performed when $\pi(u)$ contains exactly $b$ pointers.
In this case, there are too many pointers to fit in a single node.
We create a new node $v$ with \weight\ one, and evenly distribute all of the pointers and keys of $u$ and $\pi(u)$ (except for the pointer from $\pi(u)$ to $u$) between $u$ and $v$.
Now $\pi(u)$ has two children, $u$ and $v$.
The \weight\ of $u$ is set to one, and the \weight\ of $\pi(u)$ is set to zero.
As above, this does not change the relaxed depth of any leaf, so P\ref{prop-bslack-depth}$'$ still holds after Split.
Split moves a \weight\ violation from $u$ to $\pi(u)$ (closer to the root, where it can be eliminated by a Root-Zero or Root-Replace), but may create slack violations at $u$ and $v$.

\textbf{Compress.}
Compress is performed when there is a slack violation at an internal node $u$, there is no degree violation at $u$, and there are no \weight\ violations at $u$ or any of its $k \ge 2$ children.
Let $c \le kb-b$ be the number of pointers or keys stored in the children of $u$.
Compress evenly distributes the pointers or keys contained in the children of $u$ amongst the first $\lceil c/b \rceil$ children of $u$, and discards the other children.
This will also eliminate any degree violations at the children of $u$ if $c > 1$.
After the update, $u$ satisfies P\ref{prop-bslack-slack}.
Compress does not change the relaxed depth of any node, so P\ref{prop-bslack-depth}$'$ still holds after.
Compress removes at least one child of $u$, so it increases the slack of $u$ by at least one, possibly creating a slack violation at $\pi(u)$.
(However, it decreases the total amount of slack in the tree by at least $b-1$.)
Thus, after a Compress, it may be necessary to perform another Compress at $\pi(u)$.
Furthermore, as Compress distributes keys and pointers, it may move nodes with different parents together, under the same parent.
Even if two parents initially satisfied P\ref{prop-bslack-slack} (so the children of each parent contain a total of less than $b$ slack), the children of the combined parent may contain $b$ or more slack, creating a slack violation.
Therefore, after a Compress, it may also be necessary to perform Compress at some of the children of $u$.

\textbf{One-Child.}
One-Child is performed when there is a degree violation at an internal node $u$, there are no \weight\ violations at $u$ or any of its siblings, and there is no violation of any kind at $\pi(u)$.
Let $k$ be the degree of $\pi(u)$.
Since there is no slack violation at $\pi(u)$, there are a total of $c > kb-b = b(k-1)$ pointers stored in $u$ and its siblings.
Since $u$ has only one child pointer, each of its other $k-1$ siblings must contain $b$ pointers.
One-Child evenly distributes the keys and pointers of the children of $\pi(u)$. 
One-Child does not change the relaxed depth of any node, so P\ref{prop-bslack-depth}$'$ still holds after.
One-Child eliminates a degree violation at $u$, but, like Compress, it may move children with different parents together under the same parent, possibly creating slack violations at some children of $\pi(u)$.
So, it may be necessary to perform Compress at some of the children of $\pi(u)$.

All of these updates maintain P\ref{prop-rbslack-weight-zero}$'$, P\ref{prop-bslack-internal}$'$ and P\ref{prop-bslack-leaf}.
While rebalancing steps are being performed to eliminate the violation created by an insertion or deletion, there is at most one node with \weight\ zero.

We prove that a rebalancing step can be applied in any \rbslack\ that is not a \bslack. 

\begin{lem} \label{lem-relaxed-then-rebalance}
Let $T$ be a \rbslack.
If $T$ is not a \bslack, then a rebalancing step can be performed.
\end{lem}
\begin{chapscxproof}
If $T$ is not a \bslack, it contains a \weight\ violation, a slack violation or a degree violation.
If there is \weight\ violation, then Root-Zero, Absorb or Split can be performed.
Suppose there are no \weight\ violations.
Let $u$ be the node at the smallest depth that has a slack or degree violation.
Suppose $u$ has a degree violation.
If $u$ is the root, then Root-Replace can be performed.
Otherwise, $\pi(u)$ has no violation, so One-Child can be performed.
Suppose $u$ does not have a degree violation.
Then, $u$ must have a slack violation, and Compress can be performed.
%
%
\end{chapscxproof}

\section{Analysis} \label{sec-analysis}

This section provides a detailed analysis of \bslack s that store $n$ keys, by giving: an upper bound on the height of the tree, a lower bound on the average degree of nodes (and, hence, utilization), and an upper bound on the space complexity.
We first give a brief outline of the proofs and results, and then provide the full details.

Arbitrary \bslack s are difficult to analyze, so we begin by studying a class of trees called $b$-\maxslack s.
A $b$-\maxslack\ has a root with degree two, and satisfies P\ref{prop-bslack-depth}, P\ref{prop-bslack-internal} and P\ref{prop-bslack-leaf}, but instead of P\ref{prop-bslack-slack}, the children of each internal node contain a total of exactly $b$ slack.
Thus, a $b$-\maxslack\ is a \rbslack, but not a \bslack.
Consider a $b$-\maxslack\ $T$ of height $h$ that contains $n$ keys.
We prove that the total degree at depth $\delta \le h$ in $T$ is $d(\delta) = 2^{-\delta}(\alpha^{\delta}+\gamma^{\delta})$, where $\alpha = b+\sqrt{b^2-4b}$ and $\gamma = b-\sqrt{b^2-4b}$.
Since the total degree at the lowest depth is precisely the number of keys in the tree, every $b$-\maxslack\ of height $h$ contains exactly $d(h)$ keys.
Furthermore, when $h \ge 3$, we also have $(b-2)^{h} < d(h) \le b^{h}$.
Therefore, for $n > b^3$ (which implies height at least three), $h$ satisfies $\lceil \log_b n \rceil \le h \le \lceil \log_{b-2} n\rceil$.
We also prove that the average degree of nodes in $T$ is $\frac{b\cdot d(h-1)-b+2}{b\cdot d(h-2)-b+3}$, which is greater than $b-2$ for $h \ge 3$.

We next prove some connections between \maxslack s and \bslack s.
First, we show that each $b$-\maxslack\ of height $h$ has a smaller total degree of nodes at each depth than any \bslack\ of height $h$.
We do this by starting with an arbitrary \bslack\ of height $h$, and repeatedly removing pointers and keys from the children of each internal node that satisfies P\ref{prop-bslack-slack} (taking care not to violate P1, P2 or P3), until we obtain a $b$-\maxslack.
It follows that each $b$-\maxslack\ of height $h$ contains fewer keys than any \bslack\ of height $h$.
Consequently, every $b$-\maxslack\ with $n$ keys has height at least as large as any \bslack\ with $n$ keys.
We next prove that every $b$-\maxslack\ of height $h$ has a smaller average node degree than any \bslack\ of height $h$.
As above, the proof starts with an arbitrary \bslack\ of height $h$, and removes pointers and keys from nodes until the tree becomes a $b$-\maxslack.
However, in this proof, every time we remove a pointer, we must additionally show that the average degree of nodes in the tree decreases.


We then compute the \textit{space complexity} of a \bslack\ containing $n$ keys, which is the number of words needed to store it.
Consider a leaf-oriented tree with maximum degree $b$.
For simplicity, we assume that each key and each pointer to a child or data occupies one word in memory.
Thus, a leaf occupies $2b$ words, and an internal node occupies $2b-1$ words.
A memory block size of $2b$ is assumed.
Let $\bar D$ be the average degree of nodes.
Then, $U = \bar D/b$ is the proportion of space that is utilized (which we call the \textit{average space utilization} of the tree), and $1-U$ is the proportion of space that is wasted.
The space complexity is $2bF$, where $F$ is the number of nodes in the tree.
Suppose the tree contains $n$ keys.
By definition, the sum of the degrees of all nodes is $F - 1 + n$, since each node,
except the root, has a pointer into it and the degree of a leaf is the number of keys it contains.
Additionally, $F \bar D$ is equal to the sum of degrees of all nodes, so $F = (n - 1)/(\bar D - 1)$.
Therefore, the space complexity is $2b(n - 1)/(\bar D - 1)$. 
In order to compute an upper bound on the space complexity for a \bslack\ of height $h$, we simply need a lower bound on $\bar D$.
Above, we saw that $\bar D > b-2$ for \bslack s of height at least three.
It follows that a \bslack\ with $n > b^3$ keys 
has space complexity at most $2b(n-1)/(b-3) < 2n \frac{b}{b-3}$.
(Recall that $2n$ is optimal under these assumption.)
A slightly tighter upper bound is $2b(n-1) / \big( \frac{b\cdot d(h-1)-b+2}{b\cdot d(h-2)-b+3}-1 \big)$.

Section~\ref{sec-bslack-space-complexity} 
describes pathological families of B-trees, Overflow trees and H-trees, and compares the space complexity of example trees in these families with the worst-case upper bound on the space complexity of a \bslack.
By studying these families, we obtain lower bounds on the space complexity of these trees that are above the upper bound for \bslack s.

We also study the number of rebalancing steps necessary to maintain balance in a \rbslack.
Consider a \rbslack\ obtained by starting from a \bslack\ containing $n$ keys and performing a sequence of $i$ insertions and $d$ deletions.
We prove that such a \rbslack\ will be transformed back into a \bslack\ after at most $2i(2 + \lfloor \log_{\lfloor\frac{b}{2}\rfloor} (n+i)/2 \rfloor) + d/(b-1)$ rebalancing steps, irrespective of which rebalancing steps are performed, and in which order.
Hence, insertions perform amortized $O(\log(n+i))$ rebalancing steps and deletions perform an amortized constant number of rebalancing steps.

%

\subsection{Analysis of \maxslack s} \label{sec-maxslack}

For our analysis, it is helpful to generalize the definition of $b$-\maxslack s to the following.
A $(b,k)$-\maxslack\ is the same as $b$-\maxslack, except that the root has degree $k$, instead of degree two. 


We now 
compute the total degree of nodes at each depth in a $(b,k)$-\maxslack.
As we will see, for each $k$, the total degree of nodes at a given depth $\delta$ is the same in every $(b,k)$-\maxslack\ of height at least $\delta$.

\begin{lem} \label{lem-N}
The total degree of nodes at depth $\delta$ in a $(b,k)$-\maxslack\ of height $h$ is:
 \begin{displaymath}
   d(\delta, k) = \left\{
     \begin{array}{ll}
       k & \mbox{if } \delta = 0 \\
       kb-b & \mbox{if } \delta = 1 \\
       b(d(\delta-1, k) - d(\delta-2, k))& \mbox{if } 1 < \delta \le h
     \end{array}
   \right.
  \end{displaymath}
\end{lem}
\begin{chapscxproof}
In the following, we use $c(u)$ to denote the set of children of node $u$.
Let 
$T$ be a $(b,k)$-\maxslack.
The proof is by induction on $\delta$.
The base cases $\delta = 0$ and $\delta = 1$ are immediate from the definition of a $(b,k)$-\maxslack.
Consider $\delta > 1$.
Let $T_\delta$ be the set of nodes at depth $\delta$ in $T$.
Then, 
$$d(\delta,k) = \sum_{v \in T_{\delta}} deg(v) = \sum_{u \in T_{\delta-1}} \sum_{v \in c(u)} deg(v).$$
Since $T$ is an \maxslack, $$\sum_{v \in c(u)} deg(v) = deg(u)b-b = b(deg(u)-1), \ \mbox{so}$$ $$d(\delta, k) = \sum_{u \in T_{\delta-1}} b(deg(u)-1) = b(d(\delta-1,k) - |T_{\delta-1}|) = b(d(\delta-1,k) - d(\delta-2,k)).$$
\end{chapscxproof}

Since $d(\delta, k)$ is a linear homogeneous recurrence relation with constant coefficients, we use the technique described in Section~2.1.1(a) of \cite{greene1982mathematics} to obtain the following closed form solution.

\begin{lem} \label{lem-N-closed-form}
$d(\delta, k) = 2^{-\delta} (k_1 (b+\sqrt{b^2-4b})^{\delta} + k_2(b-\sqrt{b^2-4b})^{\delta}),$ where $k_1 = \frac{bk-2b}{2\sqrt{b^2-4b}} + \frac{k}{2}$ and $k_2 = k - k_1$.
\end{lem}
\details{
We obtain a closed form solution for the recurrence 
using a technique described in Section~2.1.1(a) of \cite{greene1982mathematics}.
A linear homogeneous recurrence relation with constant coefficients is an equation of the form $a_n = c_1 a_{n-1} + c_2 a_{n-2} + ... + c_d a_{n-d}$.
The coefficients $c_1, ..., c_d$ yield the characteristic polynomial $p(t) = t^d - c_1 t^{d-1} - c_2 t^{d-2} - ... - c_d$.
If the roots $r_1, ..., r_d$ of $p(t)$ are unique, then $a_n = k_1 r_1^n + k_2 r_2^n + ... + k_d r_d^n$, where the coefficients $k_1, ..., k_d$ are chosen to satisfy the base cases for the recurrence.

Observe that $d(\delta) = b(d(\delta-1) - d(\delta-2))$ can be rewritten as $a_n = b a_{n-1} + (-b) a_{n-2}$, which yields the characteristic polynomial $p(t) = t^2 - bt + b$.
The quadratic equation yields roots $\frac{b \pm \sqrt{b^2 - 4b}}{2}$, which are unique for $b > 4$.
Thus, we have $a_n = k_1 \big(\frac{b + \sqrt{b^2 - 4b}}{2}\big)^n + k_2 \big(\frac{b - \sqrt{b^2 - 4b}}{2}\big)^n$.
Plugging in the base cases, we have $a_0 = k = k_1 + k_2$ and $a_1 = kb-b = k_1 \frac{b + \sqrt{b^2 - 4b}}{2} + k_2 \frac{b - \sqrt{b^2 - 4b}}{2}$.
The first equation yields $k_2 = k - k_1$.
After some basic algebra, 
$k_1 = \frac{kb - 2b}{2\sqrt{b^2-4b}} + \frac{k}{2}$.
}

\begin{cor} \label{cor-N-closed-form-two-maxslack}
$d(\delta, 2) = 2^{-\delta} ((b+\sqrt{b^2-4b})^{\delta} + (b-\sqrt{b^2-4b})^{\delta})$.
\end{cor}
\details{
When $k=2$, $k_1 = \frac{bk-2b}{2\sqrt{b^2-4b}} + \frac{k}{2} = 1$ and $k_2 = k - k_1 = 1$.
}

Since $b+\sqrt{b^2-4b}$ asymptotically approaches $2b$ as $b$ increases, and $b-\sqrt{b^2-4b}$ approaches zero, $d(\delta, 2)$ approximately grows like $b^{\delta}$ for large $b$.
%
Simple algebra establishes the following bounds on $d(\delta, 2)$.
\begin{cor} \label{cor-N-bounds}
For $h \ge 3$, $\big(\frac{b}{2}\big)^h < d(h, 2) \le b^h.$
\end{cor}

For the following lemma, we
used symbolic mathematics software to obtain the partial derivatives of $d(\delta, k)$ with respect to $\delta$ and $k$, 
and prove that they are positive.

\begin{lem} \label{lem-N-increasing-with-delta}
$d(\delta, k)$ is an increasing function of $\delta$ and $k$.
\end{lem}

\begin{lem} \label{lem}
The total degree of nodes in every $(b,k)$-\maxslack\ of height $h$ is
\begin{displaymath}
   D(h,k) = \left\{
     \begin{array}{ll}
       k & : h = 0 \\
       k + b(d(h-1,k)-1) & : h > 0
     \end{array}
   \right.
  \end{displaymath}
Moreover, $D(h, k)$ is increasing in $h$ and $k$.
\end{lem}
\begin{chapscxproof}
By Lemma~\ref{lem-N}, $D(h,k) = \sum_{\delta=0}^h d(\delta,k) = k + (kb-b) + \sum_{\delta=2}^h b(d(\delta-1,k)-d(\delta-2,k))$.
This telescoping sum reduces to $D(h,k) = k + (kb-b) + b(d(h-1,k)-d(0,k)) = k + b(d(h-1,k)-1)$.
Lemma~\ref{lem-N-increasing-with-delta} 
implies that $D(h, k)$ is increasing in $h$ and $k$.
\end{chapscxproof}

We now consider the average degree of nodes at each depth in a $(b,2)$-\maxslack.
In any $(b,2)$-\maxslack, the two children of the root must share exactly $2b-b = b$ pointers or keys.
Let us build intuition with an example.
Consider a $(b,2)$-\maxslack\ in which the children of the root evenly share $b$ pointers or keys.
The grandchildren of the root must share a total of exactly $((b/2)b-b) + ((b/2)b-b) = b^2-2b$ pointers or keys.
Thus, the average degree of nodes at depth zero is two, at depth one is $b/2$, and at depth two is $b-2$.
We prove that every $(b,2)$-\maxslack\ $T$ of height $h$ has the smallest average node degree of any $(b,k)$-\maxslack\ of height $h$.
We first derive expressions for the average degree of nodes in an \maxslack, and the average degree at each depth.

Since the total degree of nodes at depth $\delta$ is $d(\delta,k)$, and the total number of nodes at depth $\delta$ is $d(\delta-1,k)$, we obtain the following.

\begin{lem} \label{lem-d}
The average degree of nodes at depth $\delta$ in any $(b,k)$-\maxslack\ of height $h$ is:
\begin{displaymath}
   \bar d(\delta,k) = \left\{
     \begin{array}{ll}
       k & : \delta = 0 \\
       \frac{d(\delta,k)}{d(\delta-1,k)}& : 0 < \delta \le h
     \end{array}
   \right.
  \end{displaymath}
\end{lem}

We used our mathematics software to obtain the partial derivatives of $\bar d(\delta, k)$ with respect to $\delta$ and $k$.
We proved the following two lemmas by showing that $\frac{\partial}{\partial \delta} \bar d(\delta,k) > 0$ for $2 \le k \le b-2$, $\frac{\partial}{\partial \delta} \bar d(\delta,k) < 0$ for $k > b-2$, and $\frac{\partial}{\partial k} \bar d(\delta,k) > 0$.

\begin{lem} \label{lem-d-change-with-delta}
$\bar d(\delta,k)$ is an increasing function of $\delta$ for $2 \le k \le b-2$, and is a decreasing function of $\delta$ for $k \ge b-1$.
\end{lem}
\details%
{
\begin{chapscxproof}
We prove this lemma by demonstrating that the partial derivative of $\bar d(\delta,k)$ with respect to $\delta$ is positive for $2 \le k \le b-2$, and negative for $k > b-2$.
Using our mathematics software, we obtained a very complicated rational function that we simplified to 
$$\frac{\partial}{\partial \delta} \bar d(\delta,k) = \frac{-(k^2-bk+b)(4b)^{\delta} \sqrt{b^2-4b}\big(\alpha \gamma \log_e \frac{\alpha}{\gamma}\big)}{((2k-\gamma)(\alpha^{\delta}+\gamma^{\delta}))^2},$$
where $\alpha = b + \sqrt{b^2-4b}$ and $\gamma = b - \sqrt{b^2-4b}$.
Observe that $1 < \gamma < b < \alpha < 2b$.
Thus, the denominator is positive except when $2k - \gamma = 0$.
\trevor{Proving $2k-\gamma > 0$ is MUCH easier than this... Just use $\gamma < 2$.}
This is the case precisely when $k^2 - kb + b = 0$, which occurs when $k = \frac{\gamma}{2}$ or $k = \frac{\alpha}{2}$.
Since $b > 4$, $\frac{\gamma}{2} < 2$ and $b-2 < \frac{\alpha}{2} < b-1$.
\trevor{Where the hell did I get $b-2$ and $b-1$ from? Something seems wrong.}
Therefore, the denominator is always positive when $k \ge 2$. 
Since $b \ge 5$, $4b > 0$ and $\sqrt{b^2-4b} > 0.$
Since $1 < \gamma < \alpha$, $\alpha\gamma\log_e \frac{\alpha}{\gamma} > 0$.
Therefore, $\frac{\partial}{\partial \delta} \bar d(\delta,k) > 0$ is equivalent to $k^2-bk+b < 0$.
As above, $k^2-bk+b = 0$ when $k=\frac{\gamma}{2}$ or $k=\frac{\alpha}{2}$.
For all $k \in (\frac{\gamma}{2}, \frac{\alpha}{2})$, $k^2-bk+b < 0$, so $\frac{\partial}{\partial \delta} \bar d(\delta,k) > 0$ for $2 \le k \le b-2$.
\trevor{Look at ``demonstrating that depthwise average degree is increasing with depth.nb'' to fix this proof.}
\end{chapscxproof}
}

\begin{lem} \label{lem-d-increasing-with-k}
$\bar d(\delta,k)$ is an increasing function of $k$.
\end{lem}
%

Let $T$ be a \bslack\ of height $h$.
Since every node in $T$ except for the root is pointed to by exactly one child pointer, the number of nodes in $T$ is the total degree of all nodes at depths zero through $h-1$, plus one for the root, which is exactly $D(h-1,k)+1$.
Thus, we obtain the following.

\begin{lem} \label{lem-D-closed-form}
The average degree of nodes in any $(b,k)$-\maxslack\ of height $h$ is:
 \begin{displaymath}
   \bar D(h,k) = \left\{
     \begin{array}{ll}
       k & : h = 0 \\
       \frac{D(h,k)}{D(h-1,k)+1}& : h > 0
     \end{array}
   \right.
  \end{displaymath}
\end{lem}

\begin{lem} \label{lem-D-change-with-h}
$\bar D(h,k)$ is an increasing function of $h$ for $2 \le k \le b-2$, and is a decreasing function of $h$ for $k \ge b-1$.
\end{lem}
\begin{chapscxproof}
%
The average node degree at depth $\delta \in \{0,1,...,h\}$ is the same in every $(b,k)$-\maxslack\ of height at least $h$.
By definition, $\bar D(h,k)$ must be a weighted average of the terms $\bar d(0,k), \bar d(1,k), ..., \bar d(h,k)$.
Thus, $\bar D(h+1,k)$ is a weighted average of $\bar D(h,k)$ and $\bar d(h+1,k)$
Suppose $k \le b-2$.
By Lemma~\ref{lem-d-change-with-delta}, $\bar d(h+1,k) > \max_{i \le h} \bar d(i,k)$.
Therefore, $\bar d(h+1,k) > \bar D(h,k)$, which implies that $\bar D(h+1,k) > \bar D(h,k)$.
Now, suppose $k \ge b-1$.
By Lemma~\ref{lem-d-change-with-delta}, $\bar d(h+1,k) < \min_{i \le h} \bar d(i,k)$.
Therefore, $\bar d(h+1,k) < \bar D(h,k)$, which implies that $\bar D(h+1,k) < \bar D(h,k)$.
\end{chapscxproof}

We used our mathematics software to obtain the partial derivative of $\bar D(h,k)$ with respect to $k$, and proved that it is positive, yielding the following result.

\begin{lem} \label{lem-D-increasing-with-k}
$\bar D(h,k)$ is an increasing function of $k$.
\end{lem}
\details{
We prove $\frac{\partial}{\partial k} \bar D(h,k) > 0$.
As above, we used our mathematics software to compute the derivative, yielding a function that we simplified to
$$\frac{\partial}{\partial k} \bar D(h, k) = \frac{2^h b(b-4)(2^{h+1} (1+b^{h+1}) - \alpha^{h+1} - \gamma^{h+1})}{((\gamma^h - \alpha^h)(\gamma(b+k) + 2b) + 2^{h+1}\sqrt{b^2-4b}(1+k-b))^2},$$
where $\alpha = b + \sqrt{b^2-4b}$ and $\gamma = b - \sqrt{b^2-4b}$.

We first prove that the numerator is positive.
Since $b > 4$, $2^h b(b-4) > 0$, so we need only prove 
$2^{h+1}(1+b^{h+1}) > \alpha^{h+1} + \gamma^{h+1}$.
By the binomial theorem, $(\alpha + \gamma)^{h+1} = \sum_{k=0}^{h+1} \binom{h+1}{k} \alpha^{(h+1)-k} \gamma^k = \alpha^{h+1} + \gamma^{h+1} + \sum_{k=1}^h \binom{h}{k} \alpha^{h-k} \gamma^k$.
Since $\alpha + \gamma = 2b$, $(2b)^{h+1} = \alpha^{h+1} + \gamma^{h+1} + \sum_{k=1}^h \binom{h}{k} \alpha^{h-k} \gamma^k \ge \alpha^{h+1} + \gamma^{h+1}$.
Since $2^{h+1}(1+b^{h+1}) > (2b)^{h+1}$, we have established $2^{h+1}(1+b^{h+1}) > \alpha^{h+1} + \gamma^{h+1}$. 

We now prove that the denominator is positive.
Since the denominator is of the form $(f(k))^2$, it suffices to prove $f(k) = (\gamma^h - \alpha^h)(\gamma(b+k) + 2b) + 2^{h+1}\sqrt{b^2-4b}(1+k-b) \neq 0$.
We first try to show $f(k) < 0$.
Since $\gamma < \alpha$ and $\gamma, b, k > 0$, $(\gamma^h - \alpha^h)(\gamma(b+k) + 2b) < 0$.
The expression $2^{h+1}\sqrt{b^2-4b}$ is positive, so $f(k) < 0$ will be true if $k \le b-1$.
Suppose $k = b$.
Then, $f(k) = f(b) = \alpha^h - \gamma^h - 2^{h-1} \sqrt{1-4/b}$.
We prove $f(b) > 0$.
Since $b > 4$, $f(b) > \alpha^h - \gamma^h - 2^{h-1}$.
By the binomial theorem, $2^{h-1} + \gamma^h < 2^h + \gamma^h < (2+\gamma)^h$.
Observe that $\alpha > \gamma+2$.
Thus, $2^{h-1} + \gamma^h < (2+\gamma)^h < \alpha^h$, so $f(b) > 0$.
Therefore, in all cases, $f(k) \neq 0$.
}

We proved a simple lower bound on $\bar D(h,k)$ using our mathematics software.

\begin{lem} \label{lem-D-lower-bound}
$\bar D(h,k) > b-2$ for $h \ge 3$ (and $b \ge 5$).
\end{lem}

\subsection{Relating \bslack s to \maxslack s} \label{sec-maxslack-to-bslack}

In this section, we first prove that a $(b,2)$-\maxslack\ has the greatest height of any \bslack\ containing the same number of keys, and a smaller average degree of nodes than any \bslack\ with the same height.
Then, we compute upper and lower bounds on the space used to store a \bslack\ with $n$ keys.


\begin{prop} \label{lem-internal-with-slack-has-degree-three-child}
Let $u$ be an internal node in a \bslack. 
If the total slack contained in the children of $u$ is less than $b$, then some child of $u$ has degree at least three.
\end{prop}
\begin{chapscxproof}
Suppose the total slack contained in the children $v_1, v_2, ..., v_l$ of $u$ is less than $b$.
Then, $deg(v_1)+deg(v_2)+...+deg(v_l) > lb-b$.
By P\ref{prop-bslack-internal}, $l \ge 2$.
Since $b > 4$, it follows that $lb-b = b(l-1) > 4(l-1) \ge 2l$.
Therefore, the average degree of the children of $u$ is $(deg(v_1)+deg(v_2)+...+deg(v_l))/l > \frac{lb-b}{l} > 2$.
\end{chapscxproof}

\begin{lem} \label{lem-N-worst-in-two-maxslack}
Every $(b,2)$-\maxslack\ of height $h$ has a smaller total degree of nodes, at each depth, than any \bslack\ of height $h$.
\end{lem}
\begin{chapscxproof}
Let $T$ be a \bslack\ of height $h$. 
We can transform $T$ into a $(b,2)$-\maxslack\ by removing keys and pointers.
Removing a key, or a pointer from a node that has at least three pointers, does not affect P\ref{prop-bslack-depth}, P\ref{prop-bslack-internal} or P\ref{prop-bslack-leaf}.
Let $u$ be any internal node whose children share a total of less than $b$ slack.
We arbitrarily remove a key or pointer from the child, $v$, of $u$ with the largest degree.
By Proposition~\ref{lem-internal-with-slack-has-degree-three-child}, $v$ must have degree at least three.
We can repeat this process until $T$ is a $(b,2)$-\maxslack.
\end{chapscxproof}

\begin{cor} \label{cor-2maxslack-worst-case-height}
Every $(b,2)$-\maxslack\ with $n$ keys has a larger height than any \bslack\ with $n$ keys.
\end{cor}

\begin{cor} \label{cor-2maxslack-fewest-keys}
Any \bslack\ of height $h$ contains more keys than every $(b,2)$-\maxslack\ of height $h$, and, hence, more than $d(h,2)$ keys.
\end{cor}

\begin{lem} \label{lem-D-lower-bound-for-bslack}
Every $(b,2)$-\maxslack\ of height $h$ has a smaller average node degree than any \bslack\ of height~$h$.
\end{lem}
\begin{chapscxproof}
We first describe how to transform a \bslack\ into an \maxslack\ of the same height while decreasing the average node degree.
Observe that, since an \maxslack\ satisfies P\ref{prop-bslack-depth}, P\ref{prop-bslack-internal} and P\ref{prop-bslack-leaf}, any \bslack\ will become an \maxslack\ if pointers and keys are removed until, for each internal node $u$, the children of $u$ share a total of $b$ slack.
%
The proof is by induction on the height of the tree.
Let $u$ be the root of a \bslack\ $T$. 

In the base case, the children of $u$ are leaves.
Arbitrarily removing keys from the children of $u$ until the children contain a total of exactly $b$ slack will transform $T$ into an \maxslack\ while decreasing the average degree of nodes.

Now, suppose the children of $u$ are internal.
By the inductive hypothesis, we can transform each subtree rooted at a child of $u$ into an \maxslack.
After these transformations, for every internal node in every subtree rooted at a child of $u$, the children of this internal node contain a total of $b$ slack.
If the children of $u$ contain a total of $b$ slack, then $T$ is an \maxslack.
Otherwise, we would like to remove some grandchild of $u$, to increase this slack.
As we argued in the proof of Lemma~\ref{lem-N-worst-in-two-maxslack}, removing a pointer from a node that has the largest degree amongst its siblings yields a \bslack. 
However, we must carefully choose which grandchild to remove so that we decrease the average degree of nodes.
%
Let $v$ be the child of $u$ that is the root of the tree with the largest average degree.
By Lemma~\ref{lem-D-increasing-with-k}, $v$ has the largest degree amongst its siblings.
By Lemma~\ref{lem-internal-with-slack-has-degree-three-child}, $v$ must have at least three pointers, so removing one of its children does not violate P\ref{prop-bslack-internal}.
It is easy to verify that removing one of $v$'s children will not violate P\ref{prop-bslack-depth} or P\ref{prop-bslack-leaf}.
We remove the child of $v$ that is the root of the tree with the largest average degree.
Since this tree 
has the largest average degree of any tree rooted at a child of $v$, removing it decreases the average degree of $T$. 
(This is because every other subtree rooted at a child of $v$ has the same or smaller average degree, and removing this child of $v$ decreases the degree of $v$.)
%
We can repeatedly apply this transformation until the children of $u$ contain a total of $b$ slack, at which point $T$ is an \maxslack.

We now prove the main result.
Given a \bslack\ $T$, we first transform it into an \maxslack.
Then, if the root of $T$ has more than two children, we transform $T$ into a $(b,2)$-\maxslack\ by keeping the two children 
that are the roots of the trees with the largest average degrees, and throwing away the rest.
\end{chapscxproof}

Since the average degree of nodes represents the fraction of space that is utilized, this 
implies that every $(b,2)$-\maxslack\ of height $h$ wastes a larger proportion of space than any \bslack\ of the same height.
We can also obtain a lower bound on the average degree (and, hence, the fraction of space that is utilized) for any \bslack\ containing $n$ keys.

\begin{lem} \label{lem-2maxslack-worst-case-Dbar}
A \bslack\ with $n \ge 2$ keys has average degree greater than $\bar D(\lceil \log_b n \rceil - 1, 2)$.
\end{lem}
\begin{chapscxproof}
Let $T$ be a \bslack\ of height $h$ containing $n$ keys.
By Lemma~\ref{lem-D-lower-bound-for-bslack}, $T$ has average degree greater than $\bar D(h,2)$.
Lemma~\ref{lem-D-change-with-h} implies that $\bar D(h,2)$ increases with $h$, so it suffices to find a lower bound on $h$.
Since $b$ is the maximum possible degree for any node in $T$, $h$ is at least $\lceil \log_b n \rceil - 1$.
\end{chapscxproof}

We can now compute bounds on $\func{S}(n)$, the space complexity of a \bslack\ containing $n$ keys.
Recall from Section~\ref{sec-bslack} that $\func{S}(n) = 2b(n-1)/(\bar D - 1)$.
By Lemma~\ref{lem-2maxslack-worst-case-Dbar}, $D^* \ge \bar D(\lceil \log_b n \rceil - 1, 2)$.
Let $s = \bar D(\lceil \log_b n \rceil - 1, 2)$.
Then, $$\frac{n-1}{b-1} \le F \le \frac{n-1}{s - 1} \ \mbox{ and } \ \frac{2b(n-1)}{b-1} \le \func{S}(n) \le \frac{2b(n-1)}{s - 1}.$$
%
\\

\subsection{Amortized logarithmic rebalancing} \label{sec-log-rebalancing}

%


In the following, we assume that the tree is initially a \bslack.
After a sequence of insertions and deletions, the tree is a \rbslack.
The goal of this section is to establish an upper bound on the number of rebalancing steps needed to transform this \rbslack\ back into a \bslack.

We assume that the updates shown in Figure~\ref{fig-bslack-updates} are performed sequentially.
In a concurrent setting, locks or lock-free methodologies such as the template described by Brown, Ellen and Ruppert~\cite{Brown:2014} can be used to ensure that updates appear to atomically operate on mutually exclusive sets of nodes (so that the effect will be the same as if the updates were performed sequentially in some order).

Our analysis follows the approach taken in \cite{LF95}.
Consider any arbitrary \bslack\ $T$.
Initially, we associate every key in the tree with the leaf that contains it.
When a key is inserted into a leaf $u$, we associate the key with $u$.
After a key is deleted from $u$, the key is still associated with $u$.
If the node $u$ is deleted, then all keys associated with $u$ are instead associated with another node.
Two cases arise.
If $u$ is deleted by a Root-Replace, then all keys associated with $u$ are instead associated with the only child of $u$.
Otherwise, $u$ is deleted by Absorb or Compress, and all keys associated with $u$ are instead associated with the node that was the parent of $u$ before the Absorb or Compress.

Let $\sigma$ be a sequence of updates to $T$, and $w$ be an internal node in the tree after the updates in $\sigma$ have been performed.
We define the \textit{\iset} of $w$ to be the multiset of all keys associated with nodes in the subtree rooted at $w$.
Therefore, the \iset\ of the root contains $n+i$ keys, where $i$ is the number of insertions in $\sigma$ and $n$ is the size of $T$.

We also define the \textit{relaxed height} of a node $u$ in a \rbslack.
Suppose we formed a \textit{new} \rbslack\ $T'$ by detaching the subtree rooted at $u$ from the \rbslack\ that contains it. 
The relaxed height of $u$, denoted $rh(u)$, is then the relaxed depth of the leaves in $T'$.

The following lemma relates the relaxed height of a node to the number of keys in its \iset.

\begin{lem}
Consider a \bslack\ $T$ containing at least two keys, and a sequence of updates to it.
Then, let $u$ be any node in the resulting tree.
If $u$ is the root, then its \iset\ contains at least $2 \big\lfloor \frac b 2 \big\rfloor^{rh(u)-weight(u)}$ keys.
Otherwise, its \iset\ contains at least $\big\lfloor \frac b 2 \big\rfloor^{rh(u)}$ keys.
\end{lem}
\begin{chapscxproof}
The proof is by induction on the sequence of updates performed on $T$.

\textbf{Base case.}
Let $u$ be any node in $T$, $T_u$ be the tree rooted at $u$, and $h_u$ be the height of $T_u$.
Any subtree of a \bslack\ is a \bslack, so $T_u$ is a \bslack.
By Corollary~\ref{cor-2maxslack-fewest-keys}, $T_u$ must contain at least $d(h_u,2)$ keys.
By Corollary~\ref{cor-N-bounds}, $d(h_u,2) > (\frac b 2)^{h_u}$.
Since $T_u$ is a \bslack, every node has \weight\ one, so the relaxed height of each node is equal to its height and $h_u = rh(u)$.
Therefore, $d(h_u,2) > (\frac b 2)^{rh(u)} \ge \lfloor \frac b 2 \rfloor^{rh(u)} \ge 2\lfloor \frac b 2 \rfloor^{rh(u)-1} = 2 \lfloor \frac b 2 \rfloor^{rh(u)-weight(u)}$.

\textbf{Inductive step.}
Suppose the claim holds before an update $U$.
We prove it holds after $U$.
Let $rh'(u)$ be the relaxed height of a node $u$ after $U$.

Suppose $U$ is Delete.
Then, each \iset\ remains the same, and every node has the same relaxed height before and after $U$.

Suppose $U$ is Insert.
Then, each \iset\ either gains one new key, or remains the same, and every node has the same relaxed height before and after $U$.

Suppose $U$ is Root-Replace.
Let $p$ be the old root, and $u$ be its only child.
If $u$ has \weight\ one before $U$, then $rh'(u) = rh(p)-1$.
Otherwise, $u$ has \weight\ zero before $U$, so $rh'(u) = rh(p)$.
Any keys associated with $p$ before $U$ are associated with $u$ after $U$, so the \iset\ of the root is the same before and after $U$.

Suppose $U$ is Root-Zero.
Let $r$ be the relaxed height of the root before $U$.
By the inductive hypothesis, the \iset\ of the root contains at least $2 \big\lfloor \frac b 2 \big\rfloor^{r}$ keys before $U$.
Moreover, $U$ does not change any \iset.
After $U$, the relaxed height of the root increases to $r+1$ because the \weight\ of the root changes from zero to one, so the \iset\ of the root contains at least $2 \big\lfloor \frac b 2 \big\rfloor^r = 2 \big\lfloor \frac b 2 \big\rfloor^{(r+1)-1} = 2 \big\lfloor \frac b 2 \big\rfloor^{rh(root)-weight(root)}$.

Suppose $U$ is Absorb.
Let $u$ be the child before $U$ and $p$ be its parent.
In this case, $u$ is removed by $U$, and all of its associated keys are instead associated with $p$.
The \iset, \weight\ and relaxed height of $p$ are all the same before and after $U$.

Suppose $U$ is Split.
Let $u$ be the child before $U$ and $p$ be its parent.
In this case, $U$ creates a new child, $v$, of $p$ and moves all of $p$'s pointers (except for its pointers to $u$ and $v$) into $u$ and $v$, so that $u$ and $v$ each contain at least $\lfloor \frac{b+1}{2} \rfloor \ge \frac b 2$ pointers.
Observe that $rh'(u) = rh'(v) = rh'(p) = rh(p)$.
Each pointer in $u$ or $v$ after $U$ points to a node with relaxed height $rh(p)-1$.
By the inductive hypothesis, the \iset\ of every such node contains at least $\lfloor \frac b 2 \rfloor^{rh(p)-1}$ keys.
Therefore, the \iset s of $u$ and $v$ each contain at least $\frac b 2 \lfloor \frac b 2 \rfloor^{rh(p)-1} \ge \lfloor \frac b 2 \rfloor^{rh(p)} = \lfloor \frac b 2 \rfloor^{rh'(u)} = \lfloor \frac b 2 \rfloor^{rh'(v)}$ keys, and the \iset\ of $p$ contains at least $\lfloor \frac b 2 \rfloor^{rh'(u)} + \lfloor \frac b 2 \rfloor^{rh'(v)} = 2 \lfloor \frac b 2 \rfloor^{rh'(p)}$ keys.

Suppose $U$ is Overflow.
Let $u$ be the leaf that is full.
In this case, $U$ creates a new leaf $r$ and an internal node $p$ with \weight\ zero and pointers to $u$ and $v$, and moves half of the keys from $u$ into $v$.
After $U$, the \iset\ of $p$ contains at least $b+1$ keys.
Since $u$ and $v$ are leaves, $rh(p) = 1$, so $b+1 \ge 2 \lfloor \frac b 2 \rfloor^{rh(p)}$.
The \iset s of $u$ and $v$ each contain at least $\frac b 2$ keys.
Since $rh(u) = rh(v) = 1$, $\frac b 2 \ge \lfloor \frac b 2 \rfloor^{rh(u)} = \lfloor \frac b 2 \rfloor^{rh(v)}$.

Suppose $U$ is Compress or One-Child.
Let $p$ be the upper node and $k$ be its degree.
Observe that $U$ does not change the weight or relaxed height of any node, and does not remove any key from the \iset\ of $p$.
After $U$, $p$ has $\lceil \frac c b \rceil$ children that evenly share $c$ pointers or keys.
Thus, each child contains at least $\lfloor \frac{c}{\lceil c/b \rceil} \rfloor \ge \lfloor \frac{c}{c/b+1} \rfloor = \lfloor \frac{cb}{c+b} \rfloor = \lfloor \frac{b}{1+b/c} \rfloor$ pointers or keys.
If $U$ is One-Child, then $c > kb-b$ and $k \ge 2$, so $c > b$ and $\lfloor \frac{b}{1+b/c} \rfloor > \lfloor \frac b 2 \rfloor$.
If $U$ is Compress, then two cases arise.
If $p$ has at least two children after $U$, then $\lceil \frac c b \rceil \ge 2$, so $c/b+1 \ge 2$ and $b/c \le 1$.
Therefore, each child of $p$ contains at least $\lfloor \frac{b}{1+b/c} \rfloor \ge \lfloor \frac b 2 \rfloor$ pointers or keys after $U$.
Otherwise, after $U$, the single child of $p$ contains all of the pointers and keys of the children that were removed, so its \iset\ is at least as large as it was before $U$.
The claim then follows immediately from the inductive hypothesis.
\end{chapscxproof}

\begin{cor}
Consider a \rbslack\ that results from performing a sequence of operations, $i$ of which are insertions, on a \bslack\ containing $n$ keys.
The relaxed height of the root, and, hence, any node in this \rbslack\ is at most $\big\lfloor \log_{\lfloor\frac{b}{2}\rfloor} \frac{n+i}{2} \big\rfloor + 1$.
\end{cor}

\begin{lem} \label{lem-upper-bound-on-splits}
After a sequence of operations, $i$ of which are insertions, on a \bslack\ containing $n$ keys, the total number of Absorb and Root-Zero updates that can be performed is at most $i$, and the number of Split updates that can be performed is at most $i (1+\big\lfloor \log_{\lfloor\frac{b}{2}\rfloor} \frac{n+i}{2} \big\rfloor)$.
\end{lem}
\begin{chapscxproof}
Absorb or Split is performed when a node has \weight\ zero.
Overflow is the only update that increases the number of zero \weight s in the tree, and at most $i$ Overflows occur, so there are at most $i$ nodes with \weight\ zero in the tree.
Absorb and Root-Zero each decrease the number of zero \weight s in the tree by one, so at most $i$ of these updates can be performed.
Root-Zero, Root-Replace, Absorb and Split are the only updates that can change the relaxed height of a node with \weight\ zero.
Root-Zero, Root-Replace and Absorb each change a zero \weight\ to one, and Split moves a zero \weight\ from a node with relaxed height $r$ to a node with relaxed height $r+1$.
Therefore, each zero \weight\ will remain at a node with the same relaxed height until it is moved by Split or changed to one by Root-Zero, Root-Replace or Absorb.
Since the relaxed height of any node in the tree is at most $\big\lfloor \log_{\lfloor\frac{b}{2}\rfloor} \frac{n+i}{2} \big\rfloor + 1$, each of the $i$ zero \weight s in the tree can be moved by Split at most $\big\lfloor \log_{\lfloor\frac{b}{2}\rfloor} \frac{n+i}{2} \big\rfloor + 1$ times.
\end{chapscxproof}

\begin{lem} \label{lem-amortized-rebalancing}
Let $T$ be a \rbslack\ of height $h$ that is obtained by performing any sequence of $i$ insertions and $d$ deletions on an initially empty \rbslack.
At most $2i(4+\frac 3 2 \big\lfloor \log_{\lfloor\frac{b}{2}\rfloor} \frac{n+i}{2} \big\rfloor) + 2d/(b-1)$ rebalancing steps can be applied to~$T$.
\end{lem}
\begin{chapscxproof}
Let $c$ be the total degree of the children of a parent where Compress or One-Child is performed.
We first bound the number of One-Child updates that can be performed.
If a node has exactly one pointer, we say a \textit{pointer violation} occurs at that node.
One-Child is performed only when a pointer violation occurs at a child of the parent and $c > kb-b$.
Since $c > kb-b$ and $k \ge 2$, $c \ge b+1$, so the parent will have at least two children after One-Child.
Furthermore, each child of the parent will have degree at least $\lfloor b/2 \rfloor$.
Thus, One-Child removes every pointer violation at a child of the parent, and does not create any pointer violation.
Root-Replace removes a pointer violation at the root, decreasing the number of pointer violations in the tree by one.
However, Compress can \textit{increase} the number of pointer violations in the tree by one if $c \le b$.
No other update changes the number of pointer violations in the tree.
Therefore, the total number of One-Child and Root-Replace updates that can be performed is bounded above by the number of Compress updates.

We bound the number of Compress updates by studying the change in the total amount of slack in the tree that is caused by each type of update.
It is convenient to ignore the slack in any node with a zero \weight\ value, since Compress cannot affect any such node.
Compress redistributes a total of $c < kb-b$ pointers or keys from $k$ nodes to $\lceil c/b \rceil$ nodes.
Since $c \le kb-b = b(k-1)$, $\lceil c/b \rceil \le k-1$, Compress will remove at least one node from the tree.
Removing this node removes $b$ slack, and increases slack at the parent by one.
Thus, Compress reduces the total amount of slack in the tree by at least $b-1$ (and by even more, if more than one node is removed).
Split is applied precisely when the parent of a node with \weight\ value zero contains exactly $b$ pointers (and no slack).
Since the node with \weight\ value zero contains exactly two pointers, $b+1$ pointers are moved into the nodes with \weight\ value one in Figure~\ref{fig-bslack-updates}, so Split increases the total slack in the tree by exactly $b-1$.
By Lemma~\ref{lem-upper-bound-on-splits}, Split can be performed at most $i (1+\big\lfloor \log_{\lfloor\frac{b}{2}\rfloor} \frac{n+i}{2} \big\rfloor)$ times, so the total amount of slack created by Split is at most $i(1+\big\lfloor \log_{\lfloor\frac{b}{2}\rfloor} \frac{n+i}{2} \big\rfloor)(b-1)$.
It is easy to verify that Insert, Insert-Distribute and Absorb each decrease the total slack by one, and that Delete and Insert-Overflow increase the total slack by one and $2(b-1)$, respectively.
Thus, the total slack created by Delete and Insert-Overflow updates is $d + 2i(b-1)$, so the total slack in $T$ is at most $i(1+\big\lfloor \log_{\lfloor\frac{b}{2}\rfloor} \frac{n+i}{2} \big\rfloor)(b-1) + d + 2i(b-1) = i(b-1)(3+\big\lfloor \log_{\lfloor\frac{b}{2}\rfloor} \frac{n+i}{2} \big\rfloor) + d$.
Therefore, at most $i(3+\big\lfloor \log_{\lfloor\frac{b}{2}\rfloor} \frac{n+i}{2} \big\rfloor) + d/(b-1)$ Compress updates can occur.

By Lemma~\ref{lem-upper-bound-on-splits} at most $i$ Absorb and Root-Zero updates and $i (1+\big\lfloor \log_{\lfloor\frac{b}{2}\rfloor} \frac{n+i}{2} \big\rfloor)$ Split updates can occur.
Since the total number of Root-Replace and One-Child updates is at most the number of Compress updates, the number of Root-Replace, One-Child and Compress updates that can occur is at most $2i(3+\big\lfloor \log_{\lfloor\frac{b}{2}\rfloor} \frac{n+i}{2} \big\rfloor) + 2d/(b-1)$.
Therefore, at most $2i(3+\big\lfloor \log_{\lfloor\frac{b}{2}\rfloor} \frac{n+i}{2} \big\rfloor) + 2d/(b-1) + i + i (1+\big\lfloor \log_{\lfloor\frac{b}{2}\rfloor} \frac{n+i}{2} \big\rfloor) = 2i(4+\frac 3 2 \big\lfloor \log_{\lfloor\frac{b}{2}\rfloor} \frac{n+i}{2} \big\rfloor) + 2d/(b-1)$ rebalancing steps can be applied to $T$.
\end{chapscxproof}

This result implies that the number of rebalancing steps needed to rebalance the tree after a sequence of deletions is amortized constant, and after a sequence of insertions is amortized logarithmic in: the size of the tree the last time it was a \bslack\ plus the number of insertions that have occurred since then. 
Section~\ref{sec-constant-rebalancing} explains how \bslack s can be modified to obtain amortized constant rebalancing by slightly increasing the amount of slack shared amongst the children of an internal node.

%

\section{\bslack s with amortized constant rebalancing} \label{sec-constant-rebalancing}

The main challenge in achieving amortized constant rebalancing is ensuring that long sequences of Split and Compress operations occur infrequently.
Split can necessitate other Splits higher in the tree and many Compresses.
Compress can necessitate many other Compresses.
This makes Split and Compress particularly problematic.

Split occurs only when an internal node is full.
If a Compress at an internal node leaves some slack in each of its children, 
then Splits will not immediately occur at the children.
With this in mind, we make some small modifications.
P\ref{prop-bslack-slack} is replaced with P\ref{prop-bslack-slack}$'$, which says that, for each internal node $u$ of degree $k$, the total slack contained in the children of $u$ is at most $b+k-1$ (so the worst-case slack per node is only one greater than in a standard \bslack).
A slack violation then occurs at any internal node that violates P\ref{prop-bslack-slack}$'$.
The children of an internal node of degree $k$ where a slack violation occurs will have total degree less than $kb-(b+k-1)=(k-1)(b-1)$.
Thus, Compress is performed only at internal nodes whose children have total degree $c \le (k-1)(b-1)$, and One-Child is performed only at internal nodes whose children have total degree $c > (k-1)(b-1)$.
This threshold is chosen so that Compress is only performed when it can remove one node and still leave each child with one slack (so that each child can accommodate one more key before necessitating a Split).
We then change Compress so that it evenly distributes the $c$ pointers or keys of its children amongst $\lceil \frac{c}{b-1} \rceil$ nodes, instead of $\lceil \frac{c}{b} \rceil$.
This way, each child is guaranteed to have at least one slack afterwards.

We prove that the number of rebalancing steps is amortized constant using the potential method. 
The potential of a node $u$, denoted $\phi(u)$, captures the intuition that
a node is bad if it contains too much slack, is full, or has \weight\ zero.
\begin{displaymath}
   \phi(u) = \left\{
     \begin{array}{ll}
       b-deg(u) & \mbox{if $deg(u) < b$ and $u$ has \weight\ one} \\
       b & \mbox{otherwise (i.e.,\ $deg(u) = b$ or $u$ has \weight\ zero)}
     \end{array}
   \right.
\end{displaymath}
The potential of a tree $T$, denoted $\Phi(T)$, is the sum of potentials of its nodes.

We now study how $\Phi(T)$ is changed by deletion, insertion, and each rebalancing step.
Let $u$ be a node with degree $k$. Recall that leaves never have \weight\ zero.

\textbf{Delete.}
If $u$ is full, then $\phi(u)$ changes from $b$ to $1$.
Otherwise, $\phi(u)$ changes from $b-k$ to $b-(k-1)$.
So, $\phi(u)$ increases by at most one.

\textbf{Insert.}
If the insertion fills $u$, then $\phi(u)$ changes from $1$ to $b$.
Otherwise, $\phi(u)$ changes from $b-k$ to $b-(k+1)$.
So, $\phi(u)$ increases by at most $b-1$.

\textbf{Overflow.}
A full node with potential $b$ turns into a node with \weight\ zero, which has potential $b$, and two nodes with \weight\ one that share a total of $b-1$ slack.
Thus, $b$ potential is replaced by $b+b-1=2b-1$ potential, which is an increase of $b-1$.

\textbf{Absorb.}
Let $u$ be the node with \weight\ zero. Its parent $\pi(u)$ has \weight\ one.
Beforehand, $u$ has degree two and $\pi(u)$ contains $j \leq b-1$ pointers, so is it not full.
Absorb decreases potential by $b$ by eliminating $u$.
It also moves a pointer from $u$ to $\pi(u)$.
If $j=b-1$, then $\phi(\pi(u))$ changes from 1 to $b$, increasing potential by $b-1$, for a net decrease of one.
Otherwise, $\phi(\pi(u))$ changes from $b-j$ to $b-(j-1)$, for a net decrease in potential of $b-1$.

\textbf{Split.}
Let $u$ be the node with \weight\ zero. Its parent $\pi(u)$ has \weight\ one, but it is full, so
it has potential $b$ behorehand. Afterwards, it has \weight\ zero, so its potential does not change.
Before the Split, $u$ has potential $b$, and it is split into two nodes, each of \weight\ one, that share a total of $b-1$ slack.
After the Split, the sum of their potentials is $b-1$.
Thus, the potential of the tree is decreased by one.

\textbf{Compress.}
Let $u$ be a node with $k \geq 2$ children that have total degree at most $(k-1)(b-1)$.
The modified version of Compress will leave at least one slack at each child of $u$, so the total potential of $u$'s children will be the total amount of slack they contain, which decreases by at least $b$, since
at least one of the children is removed.
Removing a child of $u$ also increases the slack at $u$ by one, which increases $\phi(u)$ by one (unless $u$ was full, in which case it decreases $\phi(u)$).
For each child of $u$ that is removed by Compress, $b$ slack is eliminated at at the children of $u$, and one slack is added at the parent.
Therefore, the total potential of the tree decreases by $b-1$ for each child removed by Compress.
Since at least one child is removed, the total potential of the tree decreases by at least $b-1$.

\textbf{One-Child.}
Since One-Child evenly distributes keys, it cannot create any more full nodes than existed beforehand.
It does not affect \weight s, and it does not remove any key or pointer.
So, One-Child does not affect the total potential of the tree.

Since $\Phi(T)$ is increased by one for Delete and $b-1$ for Insert (and Overflow), and no other operation increases it, after $i$ insertions and $d$ deletions, $\Phi(T) \le (b-1)i+d$.
We can use $\Phi(T)$ to bound the number of rebalancing steps that can be performed on $T$.
Let $C$, $A$, $S$, $R_0$, $R_r$ be the number of Compresses, Absorbs, Splits, Root-Zeros and Root-Replaces, respectively.
We immediately obtain $(b-1)C + A + S + 2(R_0 + R_r) \le \Phi(T) \le (b-1)i+d$.
Astute readers will notice that One-Child has not yet made an appearance.
By the same argument as in Lemma~\ref{lem-amortized-rebalancing}, the number of One-Childs is at most $C$.
Therefore, the number of rebalancing steps is constant per update.

In fact, we can achieve tighter bounds if we are more careful.
By the same argument as in Lemma~\ref{lem-upper-bound-on-splits}, $A + R_0 \le i$.
Additionally, the same argument used to show that the number of One-Childs is at most $C$ applies to Root-Replace, so $R_r \le C$.
Therefore, on average, there is at most one Absorb or Root-Zero per insertion, at most one Compress (and One-Child) and Root-Replace per insertion, and at most one Split per insertion.
Similarly, on average, there is at most one Absorb, Split, Root-Replace or Root-Zero per deletion, and at most one Compress (and One-Child) per $b-1$ deletions.

The increase in space complexity associated with these changes is very small.
Since the worst-case slack per node is only one greater than in a \bslack, the minimum average degree of this modified \bslack\ is at most one less than in a \bslack.
So, worst-case lower bound on utilization changes from $\bar D/b$ to $(\bar D - 1)/b$, and the space complexity upper bound changes from $2b(n-1)/(\bar D - 1) < 2b(n-1)/(b-3)$ to $2b(n-1)/((\bar D - 1) - 1) < 2b(n-1)/(b-4)$.

\section{Space complexity of competing trees} \label{sec-bslack-space-complexity}

In this section, we study the space complexity of some pathological families of B-trees, Overflow trees and H-trees.
The maximum degree of nodes is $b$, the block size is $2b$, and all trees are leaf-oriented.

\begin{itemize}
\item \textbf{B-tree.} The root has degree two, and all other nodes have degree $b/2$.
\item \textbf{Overflow tree.} The root has degree two, the internal nodes have degree $b/2$, and the leaves have degree $b-3$.
Overflow groups are chosen to be as large as possible, to minimize wasted space.
Specifically, for each parent $u$ of a leaf, $u$'s children are all in a single group, with one shared overflow node.
Thus, each overflow node is shared by $b/2$ leaves (which contain a total of $(b-3)(b/2) = b^2/2-3b/2$ keys).
\item \textbf{H-tree.} Parameters $\gamma$ and $\delta$ are chosen to be as large as possible, to minimize wasted space.
The root has degree two, the internal nodes have degree $\lceil b/\sqrt{2} \rceil$, and the leaves have degree $b-2$.
(H-trees are node-oriented, which would significantly inflate their space complexity on a system with only one block size.
The space complexity bounds shown in Figure~\ref{fig-spk} ignore this, and are thus quite charitable.
To actually achieve such good space complexity bounds for H-trees, one would have to completely redesign the data structure to be leaf-oriented.) 
\end{itemize}

We assume that a key and a pointer each occupy a single word in memory.
For each family, and each choice of maximum degree in $\{8,16,32\}$, we consider the minimum height tree from the family containing at least $10^6$ keys, and computed its space complexity.
Observe that the resulting space complexity values are \textit{lower bounds} on the worst-case space complexity for these data structures.
These space complexity values appear in Figure~\ref{fig-spk}, along with \textit{very pessimistic upper bounds} on the space complexity for any \bslack\ ($b \in \{8,16,32\}$) containing at least $50,000$ keys.
(The aforementioned upper bounds actually apply to \maxslack s, which allow the slack shared amongst the children of a node to be one greater than in a \bslack.
Additionally, despite the fact that the space complexity of \bslack s \textit{improves as the number of keys grows}, these upper bounds only assume the tree contains at least 50,000 keys, in contrast to the other data structures, which contain at least $10^6$ keys.)
Therefore, these results are actually quite charitable to the other data structures.

Nevertheless, the advantage of \bslack s is clear.
The optimal space to store $n$ keys and pointers to associated data is $2n$.
H-trees, the closest competitor to \bslack s, use more than double the space beyond what is optimal.
If these trees were modified to implement a set instead of a dictionary (by eliminating data and allowing leaves to contain up to $2b$ keys), then the optimal space would become $n$, and it is expected that relative differences in space complexity between the trees would increase further.

\renewcommand{\tabcolsep}{2mm}
\begin{figure}[tb]
\centering
\begin{tabular}{|c|ccccc|}
\hline
Max degree & B-tree & Overflow tree & H-tree & \bslack & Optimal \\
\hline
8  & $\ge 5.333n$ & $\ge 5.066n$ & $\ge 3.840n$ & $< 2.789n$ & $2.000n$ \\
16 & $\ge 4.571n$ & $\ge 3.120n$ & $\ge 2.685n$ & $< 2.301n$ & $2.000n$ \\
32 & $\ge 4.266n$ & $\ge 2.492n$ & $\ge 2.307n$ & $< 2.145n$ & $2.000n$ \\
\hline
\end{tabular}
\caption{
    Space complexity of example trees, and worst-case bound for \bslack s.
}
\label{fig-spk}
\end{figure}

\section{Experiments} \label{sec-bslack-exp}

\paragraph{Java implementation}
The \bslack\ was implemented as a sequential data structure in Java.
Each update $U$ shown in Figure~\ref{fig-bslack-updates} was implemented as follows.
A process performing $U$ creates new a node for each node on the right hand side of the depiction of $U$ in Figure~\ref{fig-bslack-updates}, and replaces the nodes on the left hand side by these new nodes.
The nodes that were replaced by $U$ are eventually reclaimed by Java's automatic garbage collection.
If a Delete, Insert or Overflow creates a violation, then the process invokes a \textbf{Cleanup} procedure to perform rebalancing steps until the tree no longer contains any violations.

For simplicity, we implemented Cleanup as a recursive procedure.
Cleanup takes the node $u$ where a violation occurs as its argument, and attempts to perform a rebalancing step to fix the violation at $u$.
If this rebalancing step creates any new violations, or replaces any nodes with existing violations and moves the violations to new nodes, then recursive invocations of Cleanup are performed to eliminate these violations.
If an invocation of Cleanup sees that the node whose violation it was supposed to fix has already been replaced, then it knows the invocation of Cleanup that replaced it will make a recursive call to fix the violation, wherever it was moved.
Thus, this invocation of Cleanup can simply return.
The downside of a recursive implementation of Cleanup is that stack overflow may occur if rebalancing steps create a large number of violations.
Other ways to implement the Cleanup procedure are discussed in Section~\ref{sec-bslack-cleanup}.

Java is not an ideal language for implementing the \bslack, since it gives very little control over memory layout.
The purpose of our Java implementation is simply to serve as a guide for any implementers who are interested in porting the \bslack\ tree to other languages.
Each node is implemented as an array of keys, and an array of child pointers.
Unlike in C/C++, in Java, arrays cannot be embedded directly in a node.
Instead, nodes contain pointers to arrays, which are located elsewhere in memory.
Thus, even after a process has loaded (a cache line that contains) a node, accessing a key or child pointer of that node still requires performing additional loads from 
potentially distant locations in main memory.
This makes the implementation somewhat inefficient.
Furthermore, the implementation does not directly satisfy the space complexity upper bounds computed in Section~\ref{sec-analysis}.
However, if we ignore these complications and pretend that keys and pointers are embedded directly inside nodes, then we can still use this implementation to study structural properties of \bslack s in practice.

\paragraph{Methodology}
We performed randomized experimental trials for several $b$ values, simulated workloads, and tree sizes.
Each trial was divided into two phases.
In the first phase, a \bslack\ was created and initialized by inserting and deleting (with 50\% probability each) keys drawn uniformly randomly from $[0,size)$, until the tree stabilized, containing approximately $size/2$ keys.
In the second phase, one million random insertions and deletions were performed with some specified probabilities, and each rebalancing step was recorded.
We considered probabilities: 50\% insertion and 50\% deletion (50i-50d), 90\% insertion and 10\% deletion (90i-10d), and 10\% insertion and 90\% deletion (10i-90d).
At the end of each trial, average degree and space complexity were computed (under the assumption that keys and pointers were actually embedded directly in nodes).

\paragraph{Results}
We discuss a small selection of the results.
For 50i-50d, $b=16$ and $size=2^{20}=$ 1,048,576, there were approximately 1.2 rebalancing steps per successful update, the average degree was approximately 15.5, and the space complexity was less than $2.209n$.
In fact, even with a rather small $size$ of $2^{12}=$ 4,096, the average degree was approximately 15.4, and the space complexity was less than $2.226n$, which is substantially better than the theoretical average degree lower bound of 12.7 and space complexity upper bound of $2.726n$.
This suggests that the performance of \bslack s is much better in practice than what is suggested by our theoretical results.
For 50i-50d, $b=32$, and $size=2^{20}$, there were approximately 1.1 rebalancing steps per successful update, the average degree was approximately 31.5, and the space complexity was less than $2.097n$.
For 10i-90d, $b=16$ and $size=2^{20}$, there was less than one rebalancing step per successful update.
Even with 90\% of updates being deletion, the average degree remained the same as in the 50i-50d case, and the space complexity was less than $2.213n$.
For 90i-10d, there were approximately 1.2 rebalancing steps per successful update.
These results suggest that little rebalancing is required for random updates on uniform keys.

\begin{figure}[t]
\centering
\includegraphics[width=0.66\linewidth]{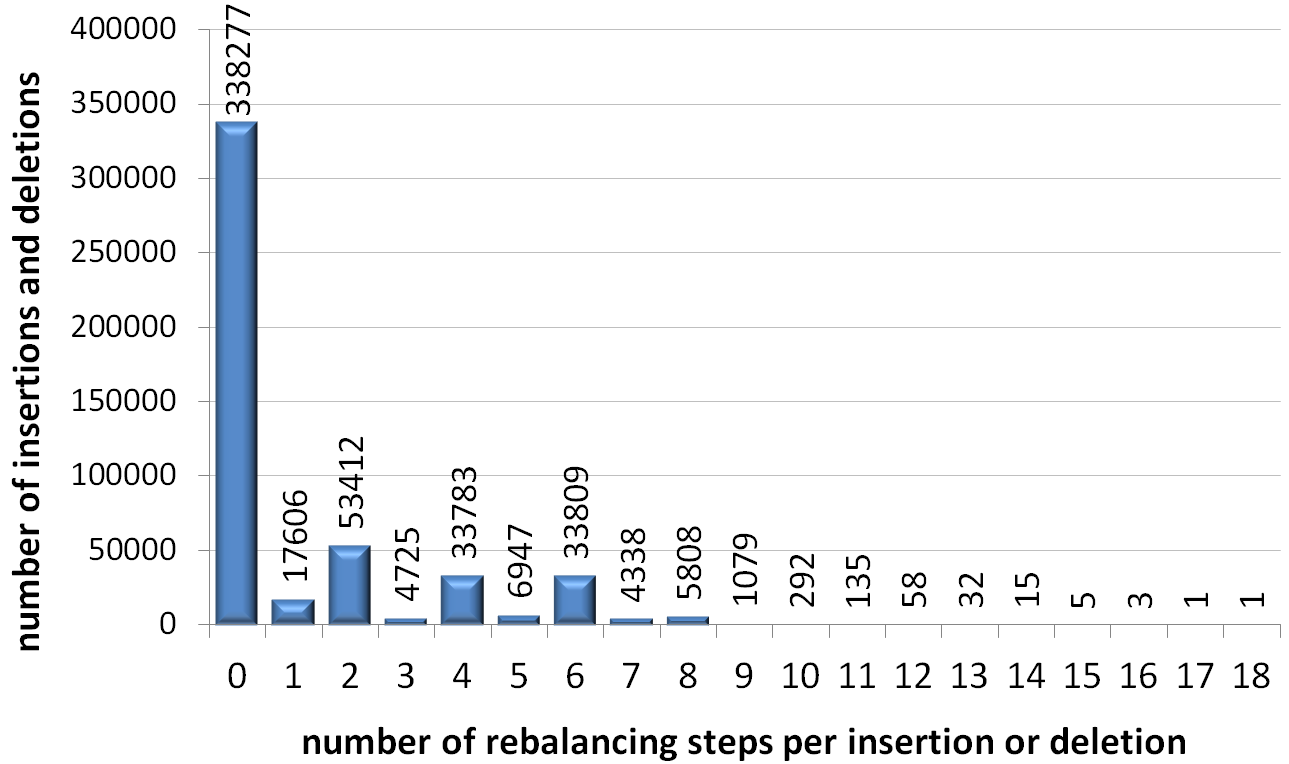}
\caption{Histogram showing the frequency of updates that perform a certain number of rebalancing steps in a randomized trial.}
\label{fig-bslack-histogram}
\end{figure}

\paragraph{Distribution of rebalancing}
It is also interesting to understand how many rebalancing steps are necessitated by each insertion or deletion. 
So, we plotted a histogram for one trial with parameters 50i-50d, $b=16$ and $size=2^{20}=$ 1,048,576 in Figure~\ref{fig-bslack-histogram}.
(Results for other trials, values of $b$ and simulated workloads are similar.)
The $x$-axis shows the number of rebalancing steps that were performed by a single insertion or deletion, and the $y$-axis shows how many insertions or deletions performed $x$ rebalancing steps.
The total number of successful insertions and deletions was 500,326, and the height of the tree was four when the trial finished.
The most rebalancing steps performed by an insertion or deletion was 18, 67.6\% of successful insertions and deletions performed no rebalancing, 97.6\% performed six or less rebalancing steps, and 99.9\% performed less than ten rebalancing steps.

\section{Implemention issues for rebalancing} \label{sec-bslack-cleanup}

One way to implement rebalancing is to explicitly maintain a collection of pointers to internal nodes where rebalancing steps must be performed.
After an update creates a violation, a rebalancing step is performed to fix that violation.
Every time a rebalancing step creates a violation, a pointer to the node where the violation occurs (and possibly also a pointer to its parent) is added to the collection.
An update does not terminate until it has emptied the queue and performed rebalancing steps to fix all violations in the tree.

Observe that violations can only occur at internal nodes.
The number of internal nodes is quite small compared to $n$ (close to $n/(b-2)^2$ in \bslack s of height at least three).
So, even if a collection contains \textit{every} internal node, the worst-case collection size may be reasonable for some applications.
Since the amortized number of rebalancing steps per update is small, most updates will result in a small collection.
We recommend using a small, fixed-size queue 
and switching to a more computationally expensive algorithm if the queue becomes full.
For instance, when a process tries to enqueue a pointer and the queue is full, it can simply discard that pointer, and continue the algorithm, recording the fact that the a pointer was discarded.
Eventually, after enough rebalancing steps are performed, the queue becomes empty, and the process can traverse the tree to find any outstanding violations, repopulate the queue, and continue the algorithm.
Although this approach is very expensive once the queue becomes full, it will not significantly increase the average running time of updates if the queue rarely becomes full.
The experimental results in Section~\ref{sec-bslack-exp} indicate that this approach could be practical, even with a very small bound on queue size.

\section{Conclusion} \label{sec-conclusion}


We introduced \bslack s, which have excellent 
space complexity in the worst case, and amortized logarithmic updates.
The data structure is simple, requires only one block size, and is well suited for concurrent implementation, both in software and in hardware.
Modifying the definition of \bslack s so that the total slack shared amongst the children of each internal node of degree $k$ is at most $b+k-1$, instead of $b-1$, yields a data structure with amortized constant rebalancing (with small constants), and only a slight increase in space complexity.
Specifically, such a tree containing $n > b^3$ keys occupies at most $\frac{2b}{b-4}n$ words.

Recently, the lock-free tree update template of Brown, Ellen and Ruppert~\cite{Brown:2014} has been used to obtain a fast concurrent implementation of \rbslack s~\cite{BrownPhD} that tolerates process crashes and guarantees some process will always make progress (Java and C++ code available at \url{http://implementations.tbrown.pro}).
In that implementation, localized updates to disjoint parts of the tree can proceed concurrently, and searches can proceed without synchronizing with updates, which makes them extremely fast.
That implementation also guarantees that, in a quiescent state, when no updates are in progress, the data structure is a (strict) \bslack.



Our sequential Java implementation of \bslack s is available at \url{http://implementations.tbrown.pro}.
Experiments have been performed to validate the theoretical worst-case bounds, and to better understand the level of pessimism in them.
The results indicate that few rebalancing steps are performed in practice, and average degree is somewhat better than the already good worst-case bounds.
For instance, for $b=16$ and $b=32$, over a variety of simulated random workloads with tree sizes varying between $2^5=$~32 and $2^{20}=$~1,048,576 keys, there were at most 1.2 rebalancing steps per insertion or deletion, and average degrees for trees were approximately $b-0.5$, which is extremely close to optimal.

\subsection*{Acknowledgments}

This work was performed while Trevor Brown was a student at the University of Toronto. Funding was provided by the Natural Sciences and Engineering Research Council of Canada. I would like thank my supervisor Faith Ellen for her helpful comments on this work, and for encouraging me to return to solutions that I had abandoned too quickly.

\bibliographystyle{abbrv}
\bibliography{bibliography}

\begin{thebibliography}{10}

\bibitem{arnow1984empirical}
D.~M. Arnow and A.~M. Tenenbaum.
\newblock An empirical comparison of \mbox{B-trees}, compact \mbox{B-trees} and
  multiway trees.
\newblock In {\em ACM SIGMOD Record}, volume 14:2, pages 33--46. ACM, 1984.

\bibitem{arnow1985p}
D.~M. Arnow, A.~M. Tenenbaum, and C.~Wu.
\newblock P-trees: Storage efficient multiway trees.
\newblock In {\em Proceedings of the 8th annual international ACM SIGIR
  conference on Research and development in information retrieval}, pages
  111--121. ACM, 1985.

\bibitem{baeza1989performance}
R.~A. Baeza-Yates and P.-A. Larson.
\newblock Performance of \mbox{B+-trees} with partial expansions.
\newblock {\em Knowledge and Data Eng., IEEE Transactions on}, 1(2):248--257,
  1989.

\bibitem{bayer1970organization}
R.~Bayer and E.~McCreight.
\newblock Organization and maintenance of large indexes.
\newblock Technical Report D1-82-0989, Boeing Scientific Research Laboratories,
  1970.

\bibitem{bronnimann2007putting}
H.~Br{\"o}nnimann, J.~Katajainen, and P.~Morin.
\newblock Putting your data structure on a diet.
\newblock {\em CPH STL Rep}, 1, 2007.

\bibitem{BrownPhD}
T.~Brown.
\newblock {\em Techniques for Constructing Efficient Lock-free Data
  Structures}.
\newblock PhD thesis, University of Toronto, 2017.
\newblock Available from \url{http://tbrown.pro}.

\bibitem{Brown:2014}
T.~Brown, F.~Ellen, and E.~Ruppert.
\newblock A general technique for non-blocking trees.
\newblock In {\em Proceedings of the 19th ACM SIGPLAN Symposium on Principles
  and Practice of Parallel Programming}, PPoPP '14, pages 329--342, 2014.
\newblock Full version available from \url{http://tbrown.pro}.

\bibitem{Browne:2000}
S.~Browne, J.~Dongarra, N.~Garner, G.~Ho, and P.~Mucci.
\newblock A portable programming interface for performance evaluation on modern
  processors.
\newblock {\em International Journal of High Performance Computing
  Applications}, 14(3):189--204, 2000.

\bibitem{culik1981dense}
K.~Culik~II, T.~Ottmann, and D.~Wood.
\newblock Dense multiway trees.
\newblock {\em ACM Transactions on Database Systems (TODS)}, 6(3):486--512,
  1981.

\bibitem{Evans:2006}
J.~Evans.
\newblock A scalable concurrent malloc (3) implementation for freebsd.
\newblock In {\em Proc. of the BSDCan Conference, Ottawa, Canada}, 2006.

\bibitem{greene1982mathematics}
D.~Greene and D.~Knuth.
\newblock {\em Mathematics for the analysis of algorithms (2nd)}.
\newblock Progress in computer science. Birkh{\"a}user, 1982.

\bibitem{Huang:1985:HTO:3857.3858}
S.-H.~S. Huang.
\newblock Height-balanced trees of order ($\beta, \gamma, \delta$).
\newblock {\em ACM Trans. Database Syst.}, 10(2):261--284, June 1985.

\bibitem{JL01abtrees}
L.~Jacobsen and K.~S. Larsen.
\newblock Variants of (a, b)-trees with relaxed balance.
\newblock {\em Int. J. Found. Comput. Sci.}, 12(4):455--478, 2001.

\bibitem{kuspert1983}
K.~K{\"u}spert.
\newblock Storage utilization in \mbox{B*-trees} with a generalized overflow
  technique.
\newblock {\em Acta Informatica}, 19(1):35--55, 1983.

\bibitem{LF95}
K.~S. Larsen and R.~Fagerberg.
\newblock B-trees with relaxed balance.
\newblock In {\em Proc.\ 9th International Symposium on Parallel Processing},
  pages 196--202, 1995.

\bibitem{relaxedbalance}
K.~S. Larsen, E.~Soisalon-Soininen, and P.~Widmayer.
\newblock Relaxed balance through standard rotations.
\newblock In F.~Dehne, A.~Rau-Chaplin, J.-R. Sack, and R.~Tamassia, editors,
  {\em Algorithms and Data Structures}, volume 1272 of {\em Lecture Notes in
  Computer Science}, pages 450--461. Springer Berlin Heidelberg, 1997.

\bibitem{Rosenberg1979}
A.~L. Rosenberg and L.~Snyder.
\newblock Compact \mbox{B-trees}.
\newblock In {\em Proceedings of the 1979 ACM SIGMOD International Conference
  on Management of Data}, SIGMOD '79, pages 43--51, New York, NY, USA, 1979.
  ACM.

\bibitem{Srinivasan01011991}
B.~Srinivasan.
\newblock An adaptive overflow technique to defer splitting in \mbox{B-trees}.
\newblock {\em The Computer Journal}, 34(5):397--405, 1991.

\end{thebibliography}

\end{document}